\documentclass{aa} % for a referee version
\usepackage{graphics}

\def\lsim{~\rlap{$<$}{\lower 1.0ex\hbox{$\sim$}}}
\def\gsim{~\rlap{$>$}{\lower 1.0ex\hbox{$\sim$}}}

\begin{document}

   \thesaurus{03     % A&A Section 6: Form. struct. and evolut. of stars
              (11.09.1 \object{Mrk~86};  % Galaxies: individual
               11.09.5;  % Galaxies: irregular,
               11.19.4;  % Galaxies: star clusters,
               11.19.5)} % Galaxies: stellar content.

   \title{Mapping the star formation history of \object{Mrk~86}$^*$}
   \subtitle{II. Stellar populations and global interpretation}

   \author{A. Gil de Paz\and J. Zamorano\and J. Gallego}

   \offprints{A. Gil de Paz\\$^*$Visiting Astronomer (AGdP), Kitt
   Peak National Observatory, National Optical Astronomy
   Observatories, which is operated by the Association of Universities
   for Research in Astronomy, Inc. (AURA) under cooperative agreement
   with the National Science Foundation.}

   \institute{Dept. de Astrof\'\i sica y CC. de la Atm\'osfera,
   Universidad Complutense de Madrid, Avda. Complutense s/n, E-28040
   Madrid\\ e-mail: gil@astrax.fis.ucm.es (AGdP)}

   \date{Received November 23, 1999; accepted July 12, 2000}

   \maketitle

   \label{firstpage}

   \begin{abstract}

In this paper, continuation of Gil de Paz et al. (Paper~I), we derive
the main properties of the stellar populations in the Blue Compact
Dwarf galaxy \object{Mrk~86}. Ages, stellar masses, metallicites and
burst strengths have been obtained using the combination of Monte
Carlo simulations, a maximum likelihood estimator and Cluster and
Principal Component Analysis. The three stellar populations detected
show well defined properties. We have studied the underlying stellar
population, which shows an age between 5-13\,Gyr and no significant
color gradients. The intermediate aged (30\,Myr old) central starburst
show a very low dust extinction with high burst strength and high
stellar mass content ($\sim$9$\times$10$^{6}$\,M$_{\sun}$). Finally,
the properties of 46 low-metallicity ($\sim$1/10\,Z$_{\sun}$)
star-forming regions were also studied.

The properties derived suggest that the most recent star-forming
activity in \object{Mrk~86} was triggered by the evolution of a
superbubble originated at the central starburst by the energy
deposition of stellar winds and supernova explosions. This superbubble
produced the blowout of a fraction of the interstellar medium at
distances of about 1\,kpc with high gas surface densities, leading to
the activation of the star formation.

Finally, different mechanisms for the star formation triggering
in this massive central starburst are studied, including the merging
with a low mass companion and the interaction with
\object{UGC~4278}. We have assumed a distance to \object{Mrk~86} of
6.9\,Mpc.

\keywords{individual: \object{Mrk~86} -- galaxies: irregular, star clusters, 
stellar content}

\end{abstract}

\section{Introduction}
\label{introduction}

The high star formation rate (SFR hereafter) and low neutral gas
content deduced for dwarf star-forming galaxies imply consumption
time-scales of about 10$^{9}$\,yr (Fanelli et al. \cite{fanelli};
Thuan \& Martin \cite{thuan81}), much shorter than the age of the
Universe. Searle et al. (\cite{searle73}) suggested that either these
objects are truly young systems or they have an intermittent star
formation history with short intense star-forming episodes followed by
long quiescent phases. Although this question remained unanswered
during decades (Thuan \cite{thuan83}; Campbell \& Terlevich
\cite{campbell}; Loose \& Thuan \cite{loose85}), nowadays, most of the
studies on dwarf star-forming systems, including Blue Compact Dwarf
galaxies (BCD hereafter), have revealed the existence of an evolved
underlying population (Kunth et al. \cite{kunth88}; Hoffman et
al. \cite{hoffman}; Papaderos et al. \cite{papaII}; Doublier
\cite{doublier98}; Norton \& Salzer \cite{norton}).

Therefore, the understanding of the mechanism, or mechanisms,
governing this star formation regulation is the stepping stone of the
dwarf star-forming galaxies evolution. Moreover, this mechanism
constitutes the missing link between the different dwarf star-forming
galaxies (Silk et al. \cite{silk87}; Burkert \cite{burkert};
Drinkwater \& Hardy \cite{drinkwater}; Papaderos et
al. \cite{papaII}).

There are two different approaches to address these questions. On the
one hand, statistical analysis of a sample of these objects will allow
the study of the relationships between fundamental parameters and
properties of the star-forming dwarf galaxies (Marlowe et
al. \cite{marlowe95}; Papaderos et al. \cite{papaI},
\cite{papaII}). On the other hand, the detailed analysis of individual
low-redshift objects with multiple regions of star formation is
fundamental to reconstruct their star formation histories. In
particular, we can obtain a better understanding of possible effects
of merging or some internal process as self-propagation that could
originate the spreading of star formation, and the effects of these
star-forming events on the future star formation.

Among dwarf galaxies, Blue Compact Dwarf galaxies conform the
subpopulation where the star-forming events are most violent. Blue
Compact Dwarfs are low-luminosity (M$_B$
$\ge$$-$18$^{\mathrm{m}}$)\footnote{For
$H_0$=75\,km\,s$^{-1}$\,Mpc$^{-1}$ (Thuan \& Martin
\cite{thuan81}).} galaxies with compact sizes whose spectra are
similar to those of low metallicity \ion{H}{ii} regions (Searle \&
Sargent \cite{searle72}; Kunth \& Sargent \cite{kunth86}; Thuan \&
Martin \cite{thuan81}). Their spectra are characterized by emission
lines over a blue continuum which implies the existence of a large
fraction of OB stars and an intense star-forming activity.

The Blue Compact Galaxy \object{Mrk~86}=\object{NGC~2537} (Shapley \&
Ames \cite{shapley32}; Markarian \cite{markarian69}), also known as
Arp~6 (Arp \cite{arp66}), constitutes an excellent laboratory to test
the BCD star formation history since its star-forming regions populate
all the galaxy. This object is the propotype of the iE galaxies, the
most important BCD galaxies subclass, that conform 70 per cent of the
BCD galaxies (Thuan \& Martin \cite{thuan81}; Thuan \cite{thuan91}).

In Gil de Paz et al. (\cite{paperI}; Paper~I hereafter) we presented
observational data for the underlying population and star-forming
regions of \object{Mrk~86}. We provided sizes, $BVRJHK$ magnitudes,
colors and emission-line fluxes for most of these regions. The
evolutionary synthesis models used in this paper were also extensively
described.

Now, we describe the procedure employed to compare the colors measured
and those predicted by the evolutionary synthesis models in
Sect.~\ref{method}. The properties of the underlying stellar
population, central starburst and recent star-forming regions are
studied in Sect.~\ref{under}, ~\ref{central} and~\ref{starforming},
respectively. Electron densities, temperatures and metal abundances
for the brightest star-forming regions in the galaxy are given in
Sect.~\ref{starformingspectra}. Finally, in
Sect.~\ref{globalinterpretation}, we provide a global interpretation
of the short- and mid-term star formation history of the galaxy. The
main conclusions from this study are summarized in
Sect.~\ref{conclusions}.

\section{Data-models comparison method}
\label{method}

\subsection{Determination of the underlying population properties}
\label{undermethod}

The optical-near-infrared colors measured for the underlying stellar
component (see Sect.~5 and 6.1 in Paper~I) have been compared with
those predicted by the Bruzual \& Charlot (priv. comm.) evolutionary
synthesis models. We have obtained the best-fitting model for each
point in the color profiles (see Fig.~6 of Paper~I) using a maximum
likelihood estimator. This maximum likelihood estimator is defined as,
\begin{equation} 
{\cal L}(t,b,Z) = \prod_{n=1}^{5} {1\over \sqrt{2\pi}\Delta C_n}
 \exp\left( - {(c_n-C_n)^2\over 2 \Delta C_n\,^2}\right)
\label{for1}
\end{equation}
where $C_n$ (with $n$=1-5) are the $B-V$, $V-R$, $R-J$, $J-H$ and
$R-K$ colors measured, and $c_n$ are those predicted by the
evolutionary synthesis models. The colors measured were corrected for
Galactic extinction using an extinction in the $B$-band of
0.15$^{\mathrm{m}}$ (Burstein \& Heiles \cite{burstein}).

We have studied different star formation histories for the
formation of this component. In particular, we have considered
instantaneous and 1, 3 and 7\,Gyr duration bursts and continuous star
formation models. This comparison has been restricted to models with
metallicity lower than the solar value. We are confident with this
assumption since the gas metallicities derived in
Sect.~\ref{starformingspectra} for the galaxy star-forming regions are
lower than one tenth solar.

\subsection{Determination of the star-forming regions properties}
\label{formingmethod}

A more elaborated comparison method has been used in the case of the
galaxy star-forming regions. This comparison method is fully described
in Gil de Paz et al. (\cite{nirucm}). Briefly, it combines Monte Carlo
simulations and a maximum likelihood estimator with Cluster and
Principal Component Analysis.

The maximum likelihood estimator employed is very similar to that
described in Sect.~\ref{undermethod}, but replacing the $R-J$ and
$R-K$ colors by the $V-J$ and $J-K$ colors. In addition, since these
regions have intense H$\alpha$ emission, we have included a new term,
defined as $m_R$+2.5$\times$log($F_{\mathrm{H}\alpha}$). This term is
equivalent to the H$\alpha$ equivalent width (EW hereafter) term,
2.5$\times$log\,EW(H$\alpha$), used in Gil de Paz et
al. (\cite{nirucm}). The $m_R$ magnitudes are those measured within
the apertures given in Paper~I.

\begin{table*}
\caption[]{Age and mass-to-light ratio for the underlying stellar population. The best-fitting model in all these solutions is two-fifths solar metal abundant. Solutions for an instantaneous, 1, 3 and 7\,Gyr duration burst and constant star formation rate models are shown.}
\begin{tabular}{c|cccc|cccc|cccc}
\hline
 & \multicolumn{2}{c}{$A_{B,\mathrm{i}}$$^{\dagger}$=0.0$^{\mathrm{m}}$} & \multicolumn{2}{c}{$A_{B,\mathrm{i}}$=0.20$^{\mathrm{m}}$} & \multicolumn{2}{c}{$A_{B,\mathrm{i}}$=0.0$^{\mathrm{m}}$} & \multicolumn{2}{c}{$A_{B,\mathrm{i}}$=0.20$^{\mathrm{m}}$} & \multicolumn{2}{c}{$A_{B,\mathrm{i}}$=0.0$^{\mathrm{m}}$} & \multicolumn{2}{c}{$A_{B,\mathrm{i}}$=0.20$^{\mathrm{m}}$} \\
\hline
 $d$ & $t$ &        $M/L_{K}$$^{\ddagger}$    & $t$ &       $M/L_{K}$   & $t$ &        $M/L_{K}$      & $t$ &       $M/L_{K}$   & $t$ &        $M/L_{K}$      & $t$ &       $M/L_{K}$\\ 
(kpc)&(Gyr)& &(Gyr)& &(Gyr)& &(Gyr)& &(Gyr)& &(Gyr)& \\ 
\hline
 & \multicolumn{4}{c}{{\bf Instantaneous burst}} & \multicolumn{4}{c}{{\bf 1\,Gyr burst}}& \multicolumn{4}{c}{{\bf 3\,Gyr burst}}\\
\hline
0.88 & 5.0 & 0.62 & 2.2 & 0.41 & 5.5 & 0.62 & 3.0 & 0.44 & 6.5 & 0.62 & 4.0 & 0.43  \\ 
0.97 & 5.2 & 0.63 & 2.4 & 0.44 & 6.0 & 0.64 & 3.2 & 0.46 & 7.2 & 0.67 & 4.2 & 0.45  \\
1.06 & 6.0 & 0.68 & 3.8 & 0.56 & 6.5 & 0.68 & 3.8 & 0.48 & 8.5 & 0.74 & 5.0 & 0.51  \\
1.16 & 8.7 & 0.83 & 5.2 & 0.63 & 9.8 & 0.86 & 5.8 & 0.63 & 10.5 & 0.85 & 6.5 & 0.62 \\
1.27 & 8.0 & 0.80 & 3.8 & 0.56 & 8.5 & 0.80 & 5.2 & 0.61 & 9.2 & 0.79 & 5.8 & 0.56  \\
1.36 & 9.5 & 0.87 & 5.5 & 0.64 & 9.8 & 0.86 & 6.0 & 0.64 & 11.2 & 0.88 & 7.2 & 0.67 \\
1.47 & 9.5 & 0.87 & 5.8 & 0.66 & 9.8 & 0.86 & 6.2 & 0.66 & 11.2 & 0.88 & 7.2 & 0.67 \\
1.57 & 9.5 & 0.87 & 5.5 & 0.64 & 9.8 & 0.86 & 6.0 & 0.64 & 11.2 & 0.88 & 7.2 & 0.67 \\
\hline				 
\end{tabular}
\label{underresultstable}
\vspace{0.3cm}

\begin{tabular}{c|cccc|cccc}
\hline
& \multicolumn{2}{c}{$A_{B,\mathrm{i}}$=0.0$^{\mathrm{m}}$} & \multicolumn{2}{c}{$A_{B,\mathrm{i}}$=0.20$^{\mathrm{m}}$} & \multicolumn{2}{c}{$A_{B,\mathrm{i}}$=0.0$^{\mathrm{m}}$} & \multicolumn{2}{c}{$A_{B,\mathrm{i}}$=0.20$^{\mathrm{m}}$} \\
\hline
 $d$ & $t$ &        $M/L_{K}$      & $t$ &       $M/L_{K}$   & $t$ &        $M/L_{K}$      & $t$ &       $M/L_{K}$   \\ 
(kpc)&(Gyr)& &(Gyr)& &(Gyr)& &(Gyr)& \\ 
\hline
 & \multicolumn{4}{c}{{\bf 7\,Gyr burst}}& \multicolumn{4}{c}{{\bf Continuous}}\\
\hline
0.88 & 8.5 & 0.58 & 7.2 & 0.49  & $>$20 & $>$0.73 & 19.2  & 0.72    \\
0.97 & 9.2 & 0.64 & 7.5 & 0.51  & $>$20 & $>$0.73 & $>$20 & $>$0.73 \\
1.06 & 10.8 & 0.74 & 7.8 & 0.53 & $>$20 & $>$0.73 & $>$20 & $>$0.73 \\
1.16 & 12.2 & 0.82 & 9.0 & 0.62 & $>$20 & $>$0.73 & 20.0  & 0.73    \\
1.27 & 11.2 & 0.76 & 7.8 & 0.53 & $>$20 & $>$0.73 & $>$20 & $>$0.73 \\
1.36 & 13.0 & 0.86 & 9.2 & 0.64 & $>$20 & $>$0.73 & 20.0  & 0.73    \\
1.47 & 13.8 & 0.90 & 9.8 & 0.67 & $>$20 & $>$0.73 & 20.0  & 0.73    \\
1.57 & 13.0 & 0.86 & 9.2 & 0.64 & $>$20 & $>$0.73 & 20.0  & 0.73    \\
\hline
\end{tabular}
\\
$^{\dagger}$ $A_{B,\mathrm{i}} \equiv A_{B,\mathrm{internal}}$\\
$^{\ddagger}$ $M/L_{K}$ is expressed in M$_{\sun}$/L$_{K,\sun}$\\
\end{table*}

In order to properly derive this new term, we have computed the
fraction of H$\alpha$ flux, i.e. the fraction of Lyman photons, due to
the stellar continuum measured within the apertures. Two different
approaches can be followed. First, we could measure the H$\alpha$
fluxes using these apertures. However, since the H$\alpha$ emission is
usually more extended than the continuum emisson, this procedure would
sistematically underestimate the H$\alpha$ flux (see, e.g. \#8, \#13,
\#18, \#50, \#70 and \#80 regions). Therefore, we have used an
alternative method. We measured the total H$\alpha$ using the {\sc
cobra} program (see Paper~I). Then, we assumed that the fraction of
photons emitted within the apertures relative to the total emission is
equivalent for the Lyman and $R$-band continuum. Thus, considering
that the apertures were obtained at e-, e$^{2}$- or e$^{3}$-folding
radii and assuming gaussian ligth profiles, this light fraction can be
computed for each region. The values obtained for this fraction, $f$,
are given in Table~\ref{tableres}.

Then, multiplying these flux ratios by the total H$\alpha$ fluxes
given in Table~4 of Paper~I, we derive the H$\alpha$ luminosities due
to the continuum emission measured within the apertures.

The H$\alpha$ fluxes were corrected for extinction using the
$E(B-V)_{\mathrm{gas}}$ color excesses provided by the
H$\beta$-H$\alpha$, H$\gamma$-H$\beta$ Balmer decrements. In addition,
the broad-band magnitudes and colors were corrected for extinction
assuming that the extinction affecting the stellar continuum and the
gas extinction are related {\it via}
$E(B-V)_{\mathrm{continuum}}$=0.44$\times$$E(B-V)_{\mathrm{gas}}$
(Calzetti et al. \cite{calzetti}). In those regions where Balmer line
ratios were not measurable we assumed an average extinction of
$E(B-V)_{\mathrm{gas}}$=0.34$^{\mathrm{m}}$. This value was obtained
as the mean of the color excesses given in Table~5 of Paper~I for
those regions with accessible Balmer line ratios.

Thus, each star-forming region has a point associated in the $B-V$,
$V-R$, $V-J$, $J-H$, $J-K$,
$m_R$$+$2.5$\times$log($F_{\mathrm{H}\alpha}$) six-dimensional
space. However, the corresponding uncertainties transform these points
into probability distributions. Using a Monte Carlo method with
10$^{3}$ points and assuming gaussian errors we reconstructed these
probability distributions. Then, we compared each of these 10$^{3}$
points with our models using the maximuum likelihood estimator
described above. Since these models are parametrized in age, $t$;
burst strength, $b$; and metallicity, $Z$, of the burst stellar
population, this method effectively provides the ($t$,$b$,$Z$)
probability distribution for each input region (see
Table~\ref{tableres}).

Finally, we studied the clustering pattern present in these
distributions using a hierarchical clustering method (see Murtagh \&
Heck \cite{murtagh}). This method allows to isolate different
solutions in the ($t$,$b$,$Z$) space. We grouped the 10$^{3}$
($t$,$b$,$Z$) points in three clusters of solutions. Then, we
performed a Principal Component Analysis (see Morrison
\cite{morrison}) for each individual solution (see Gil de Paz et
al. \cite{nirucm} for a more complete description of this procedure).

For the central starburst component we used a similar
procedure. However, since no H$\alpha$ emission was detected for this
component, the $m_R$$+$2.5$\times$log($F_{\mathrm{H}\alpha}$) term was
not included in the maximum likelihood estimator. In addition, we
introduced the continuum color excess $E(B-V)$ as a free
parameter. Color excesses in the range 0.0-1.0$^{\mathrm{m}}$ were
studied, where 0.05$^{\mathrm{m}}$ is the Galactic color excess. For
each of the 10$^{3}$ Monte Carlo particles the full range was
explored, obtaining the best-fitting color excess and ($t$,$b$,$Z$)
array.

\section{Results}
\label{results}

\subsection{Underlying population}
\label{under}

In this section we describe the results obtained after comparing the
colors measured and those predicted by the evolutionary synthesis
models.

We compared the model predictions with the optical-near-infrared
colors measured without correcting for internal extinction
($A_{B,\mathrm{internal}}$=0.0$^{\mathrm{m}}$) and using an extinction
correction factor of 0.20$^{\mathrm{m}}$ in the $B$-band. The latter
value corresponds approximately to two times the extinction of a
Galactic-type disk with an inclination of 40$^{\circ}$ relative to the
plane of the sky (see Gil de Paz et al. \cite{GZG}, GZG hereafter; see
also Gil de Paz \cite{gilphd}). Therefore, the results obtained
applying these extinction correction factors can be taken as upper and
lower limits for the age and mass-to-light ratio of this population,
respectively.

The same metallicity, 2/5\,Z$_{\sun}$, was obtained at any
galactocentric distance in the interval 0.9-1.6\,kpc. Since
these results, basically age and mass-to-light ratio, were obtained
using only aperture colors, they are not affected by the distance
uncertainty described in Sect.~2.1 of Paper~I.

The ages derived range between 5.0\,Gyr at galactocentric distances of
about 0.9\,kpc and 9.5\,Gyr at distances larger than 1.3\,kpc (for an
instantaneous burst and $A_{B,\mathrm{internal}}$=0.0$^{\mathrm{m}}$). The
corresponding interval in $K$-band mass-to-light ratio was
0.62-0.87\,M$_{\sun}$/L$_{K,\sun}$ (see
Table~\ref{underresultstable}). The change in the age and
mass-to-light ratio of the stellar population at distances shorter
than 1.3\,kpc is due to contamination from the {\it
plateau} component (see the $B$-band profile decomposition given in
Papaderos et al. \cite{papaI}). Since the spatial extent of this
component coincides with the H$\alpha$ emitting zone (see Fig.~7 in
Paper~I), this contamination is probably related with a progressively
higher contribution of the recent star-forming regions to the total
emission.

The outer region of the galaxy color profiles seems to indicate that
no significant age gradients are present. However, a small positive
metallicity or extinction gradient could compensate the existence of a
negative age gradient, or {\it vice} {\it versa}, reproducing the
observed color profiles.

The results shown in Table~\ref{underresultstable} also suggest
that, although the age of the underlying stellar population
($d$$>$1.3\,kpc) could range between 5 and 13\,Gyr depending on the
star formation history considered, the mass-to-light ratio in the
$K$-band is very well constrained for a given extinction
correction. Thus, the $K$-band mass-to-light ratio for
$A_{B,\mathrm{internal}}$=0.0$^{\mathrm{m}}$ is approximately 0.87\,$M_{\sun}/L_{K,\sun}$ and
0.65\,$M_{\sun}/L_{K,\sun}$ for $A_{B,\mathrm{internal}}$=0.20$^{\mathrm{m}}$.

The small differences ($<$30~per cent) obtained in the maximuum
likelihood estimator after comparing our data with models with burst
duration shorter than 7\,Gyr prevent us to infer the star formation
history and internal extinction of the underlying stellar
population. Therefore, we are not able to determine if this stellar
population has effectively formed in a instantaneous burst or during
long (several Gyr) periods of time as it has been observed in
\object{I~Zw~18} (Aloisi et al. \cite{aloisi99}). Only 
continuous star formation models (or with a duration for the burst
longer than 7\,Gyr) can be rule out since their very low maximuum
likelihood estimators and too old ages derived.

The evolutionary synthesis models developed for the analysis of the
star-forming regions properties only depend on the observed colors and
mass-to-light ratio of the underlying stellar population. Therefore,
the conclusions given in Sect.~\ref{starforming} for the study of
these regions are not affected by our ignorance on the past star
formation history of the galaxy.

In order to build these models (see Paper~I and
Sect.~\ref{starforming}) we adopted a mass-to-light ratio for the
underlying stellar population of 0.87\,M$_{\sun}$/L$_{K,\sun}$ in the
$K$-band, which corresponds to a null internal extinction
value. However, if the internal extinction was relatively higher,
e.g. $A_{B,\mathrm{internal}}$=0.20$^{\mathrm{m}}$, the mass-to-light
ratio could be a 25~per cent lower yielding slightly different
properties for the most recent star-forming regions (see
Sect.~\ref{starforming}).

\subsection{Central starburst}
\label{central}

After applying the comparison procedure described in
Sect.~\ref{formingmethod}, we obtained the three clusters of
solutions in the age, burst strength, metallicity and color excess
four-dimensional space. Two of these three solutions show probabilites
lower than 1~per cent. In Fig.~\ref{centralfig} we show the
distribution of the total number of solutions obtained within the
remaining solution cluster which has a probability of 98~per cent.

\begin{figure}
%\resizebox{7cm}{!}{\includegraphics{fig1.eps}} 
\vspace{14cm}
\caption{Frequency histograms for the central starburst {\bf a)} age, {\bf b)} burst strength, {\bf c)} continuum color excess and {\bf d)} stellar mass.}
\label{centralfig}
\end{figure}

The age obtained for the starburst component is about 30\,Myr, and the
burst strength is 20~per cent. Fig.~\ref{centralfig}c shows that the
continuum color excess is very well constrained between 0.06 and
0.08$^{\mathrm{m}}$. The starburst age derived agrees with the absence
of H$\alpha$ emission for this component. The expected H$\alpha$
equivalent width in emission at ages older than 30\,Myr and 20~per
cent burst strength is lower than 2\,\AA\ for any stellar
metallicity\footnote{The comparison of the integrated colors with pure
burst models results in a intermediate age ($\sim$1\,Gyr), like that
used in GZG. However, the use of the predictions of evolutionary
synthesis models for composite stellar populations yields younger
ages. Since the contribution of the central starburst component to
the total mass distribution given in GZG is negligible, the
conclusions derived in that paper are not affected by this new, more
reliable, age determination.}.

\begin{table}
\caption[]{Spectroscopic indexes for the underlying population (2/5\,Z$_{\sun}$ metal abundant), starburst component (2/5\,Z$_{\sun}$ and 1/5\,Z$_{\sun}$ metal abundant) and for the composite stellar population using a 30\,Myr old burst with burst strengths 20, 10 and 5~per cent. EW(H$\delta$) and Fe5270 and Fe5406 indexes are expressed in \AA.}
\begin{tabular}{lccccc}
\hline
\multicolumn{6}{c}{\bf Starburst metallicity: Z=2/5\,Z$_{\sun}$}\\
              &     Under.      &    Starb.         &  Composed  & Obs.      & Cor.\\
\hline  
              &                 &                   & b=0.2/0.1/0.05  &      &           \\
\hline
Mg2           & 0.186            & 0.042            & 0.05/0.07/0.08  &  0.06 & 0.08 \\
EW$_{\mathrm{H}\delta}$ & 2.91   & 6.18             & 5.89/5.59/5.14  &  6.0 & \\
D$_{4000}$    & 2.01             & 1.14             & 1.19/1.24/1.32  &  1.38 & 1.38 \\
Fe5270        & 2.508            & 0.625            & 0.95/1.23/1.57  &  1.20 & 1.10 \\
Fe5406        & 1.369            & 0.371            & 0.54/0.69/0.87  &  0.74 & 0.66 \\
\hline
\multicolumn{6}{c}{\bf Starburst metallicity: Z=1/5\,Z$_{\sun}$}\\
              &     Under.      &    Starb.         &  Composed  & Obs.      & Cor.\\
\hline  
              &                  &                  & b=0.2/0.1/0.05  &      &           \\
\hline
Mg2           & 0.186            & 0.035            & 0.05/0.06/0.08  &  0.06     & 0.08 \\
EW$_{\mathrm{H}\delta}$ & 2.91   & 6.43             & 6.11/5.78/5.30  &  6.0     &  \\
D$_{4000}$    & 2.01             & 1.16             & 1.21/1.26/1.34  &  1.38     & 1.38 \\
Fe5270        & 2.508            & 0.723            & 1.03/1.29/1.60  &  1.20     & 1.10 \\
Fe5406        & 1.369            & 0.374            & 0.54/0.69/0.86  &  0.74     & 0.66 \\
\hline
\end{tabular}
\label{indexestable}
\end{table}

In order to confirm these results, we will compare the H$\delta$
equivalent width and D$_{4000}$, Mg2, Fe5270 and Fe5406 spectroscopic
indexes measured with the values predicted by the evolutionary
synthesis models. Unfortunately, we only dispose of spectroscopic
index predictions for the case of pure burst models. Therefore, we
will derive the spectroscopic indexes of this component using
the predictions from pure burst models and the burst strength given
above. In this way, a molecular index like Mg2 (see Gorgas et
al. \cite{gorgas93}), can be written for a composite stellar
population as
\begin{eqnarray}
\mathrm{Mg2} = \mathrm{Mg2}_{\mathrm{s}} - 2.5 \log \frac{1 + \frac{1-b}{b}\frac{MLR_{\mathrm{s}}}{MLR_{\mathrm{u}}} 10^{-0.4 (\mathrm{Mg2}_{\mathrm{u}} - \mathrm{Mg2}_{\mathrm{s}})}} {1+ \frac{1-b}{b}\frac{MLR_{\mathrm{s}}}{MLR_{\mathrm{u}}}}   
\end{eqnarray}
where, Mg2$_{\mathrm{u}}$ and Mg2$_{\mathrm{s}}$ are the spectroscopic
indexes for the underlying stellar population and the young starburst,
$MLR_{\mathrm{u}}$ and $MLR_{\mathrm{s}}$ are the mass-to-ligth ratios
at the continuum and $b$ is the burst strength in mass. The
mass-to-ligth ratios $MLR_{\mathrm{u}}$ and $MLR_{\mathrm{s}}$ are,
respectively, 3.436 and 0.084\,M$_{\sun}$/L$_{B,\sun}$ in the $B$-band
(Bruzual \& Charlot priv. comm. for Z=2/5\,Z$_{\sun}$), using
$M_{B,\sun}$=5.51 (Worthey \cite{worthey}). In the case of an atomic
index (Fe5270 and Fe5406) or the equivalent width of H$\delta$, it can
be derived using
\begin{equation}
\mathrm{EW(H}\delta\mathrm{)} = \frac{(1-b) MLR_{\mathrm{s}} \mathrm{EW(H}\delta\mathrm{)}_{\mathrm{u}} +
b\ MLR_{\mathrm{u}} \mathrm{EW(H}\delta\mathrm{)}_{\mathrm{s}}}{(1-b)
MLR_{\mathrm{s}} + b\ MLR_{\mathrm{u}}}
\end{equation}
where EW(H$\delta$)$_{\mathrm{u}}$ and EW(H$\delta$)$_{\mathrm{s}}$
are the H$\delta$ equivalent widths (or Fe5270 and Fe5406 indexes) of
the underlying and young stellar populations.
  
Finally, we define the D$_{4000}$ index (Bruzual \cite{bruzual}; Gorgas et
al. \cite{gorgas}) as
\begin{equation}
\mathrm{D}_{4000}=\frac{\int_{4250 (1+z)}^{4050 (1+z)} \lambda^2 f_{\lambda} d\lambda}{\int_{3750 (1+z)}^{3950 (1+z)} \lambda^2 f_{\lambda} d\lambda}
\end{equation}
Then, assuming 
\begin{equation}
\frac{\int_{4250 (1+z)}^{4050 (1+z)} \lambda^2 f_{\lambda,\mathrm{u}} d\lambda}{\int_{4250 (1+z)}^{4050 (1+z)} \lambda^2 f_{\lambda,\mathrm{s}} d\lambda}= \frac{1-b}{b}\frac{MLR_\mathrm{s}}{MLR_\mathrm{u}}
\end{equation}
where $f_{\lambda,\mathrm{u}}$ and $f_{\lambda,\mathrm{s}}$ are the
fluxes per unit wavelength of the underlying and starburst populations
($f_{\lambda}$=$f_{\lambda,\mathrm{u}}$$+$$f_{\lambda,\mathrm{s}}$), we obtain
the following expression for the D$_{4000}$ index,
\begin{equation}
\mathrm{D}_{4000}=\frac{(1-b) MLR_\mathrm{s} + b\ MLR_\mathrm{u}}{(1-b) MLR_\mathrm{s} (\frac{1}{\mathrm{D}_{4000,\mathrm{u}}}) 
+ b\ MLR_\mathrm{u} (\frac{1}{\mathrm{D}_{4000,\mathrm{s}}}) }
\end{equation}
The continuum mass-to-ligth ratios used were those predicted for the
$B$-band in the Mg2, EW(H$\delta$) and D$_{4000}$ cases and for the
$V$-band in the case of the iron indexes
($MLR_{\mathrm{u},V}$=2.80\,M$_{\sun}$/L$_{V,\sun}$ and
$MLR_{\mathrm{s},V}$=0.0147\,M$_{\sun}$/L$_{V,\sun}$ for
Z=2/5\,Z$_{\sun}$). The latter mass-to-light ratios were obtained using
$M_{V,\sun}$=4.84 (Worthey \cite{worthey}).

\begin{figure}
\resizebox{\hsize}{!}{\includegraphics*[110,40][468,582]{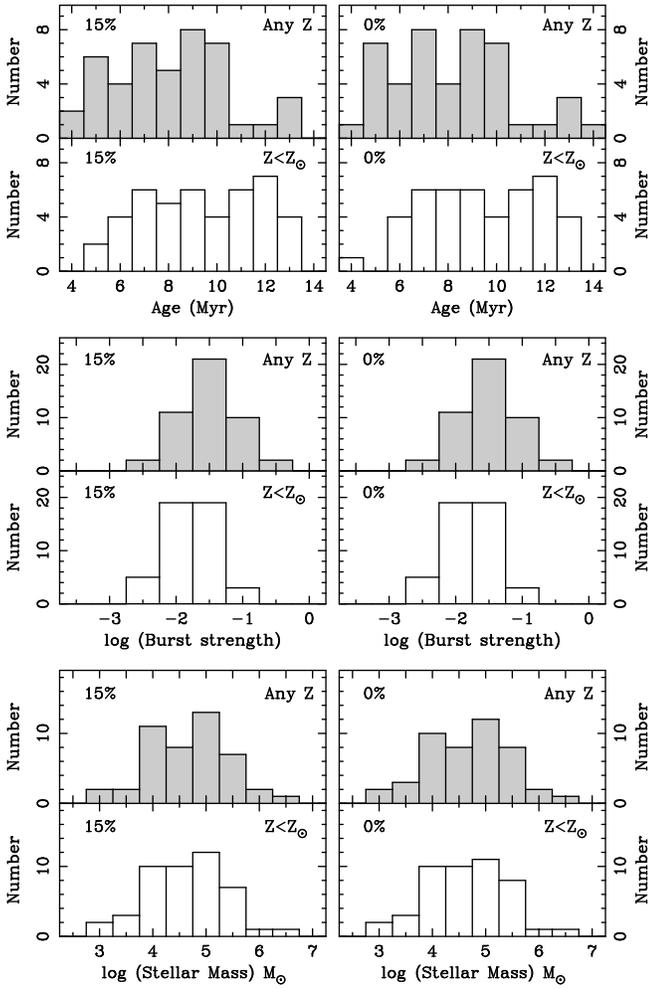}} 
\caption{Frequency histograms for the age (upper panels), burst strength (central panels) and stellar mass (lower panels) of the star-forming regions. These distributions have been obtained considering only one solution for each star-forming region. This solution corresponds to the mean of the highest-probability solution cluster. (see Table~\ref{tableres})}
\label{multhistos}
\end{figure}

\begin{table*}
\caption[]{Mean value and standard deviation for the age, burst 
strength, mass and metallicity of each individual cluster of
solutions. Only the properties of those clusters with probability
higher than 20~per cent for each region in column~1 are
given. Results for models with 15~per cent and a null fraction of Ly
photons escaping from the galaxy are shown separately. Probabilities
for these clusters of solutions are given in columns 6 and 11.
Metallicity is expressed as log($Z$/$Z_{\sun}$).}
\begin{tiny}
\begin{bf}
\begin{tabular}{r|ccccc|ccccc|c}
\hline
\multicolumn{12}{c}{Any metallicity}\\
\hline
& \multicolumn{5}{|c|}{15\% escaping Ly photons} & \multicolumn{5}{|c|}{0\% escaping Ly photons} & \\
 \# & Age & log $b$ & $Z$ &   log $M$  & Probab. & Age & log $b$ & $Z$ &   log $M$  & Probab. & $f$ \\
    &(Myr)&         &     &(M$_{\sun}$)&    (\%)     &(Myr)&         &     &(M$_{\sun}$)&    (\%)     &(\%) \\
\hline
6 &  6.03$\pm$0.73 & $-$2.02$\pm$0.34 & $-$0.45$\pm$0.11 &  3.76$\pm$0.35 & 94.9 &   6.00$\pm$0.75 & $-$2.03$\pm$0.35 & $-$0.45$\pm$0.12 &  3.75$\pm$0.35 & 94.6 & 72.0\\
7 &  6.76$\pm$1.06 & $-$2.14$\pm$0.34 &  0.31$\pm$0.17 &  3.94$\pm$0.35 & 76.7 &   6.75$\pm$1.05 & $-$2.14$\pm$0.34 &  0.31$\pm$0.16 &  3.94$\pm$0.35 & 76.5 & 64.2\\
7 &  6.98$\pm$0.69 & $-$2.47$\pm$0.22 & $-$0.41$\pm$0.06 &  3.72$\pm$0.22 & 22.7 &   6.98$\pm$0.68 & $-$2.47$\pm$0.22 & $-$0.41$\pm$0.06 &  3.72$\pm$0.22 & 22.9 &     \\
8 &  5.57$\pm$0.43 & $-$1.75$\pm$0.17 &  0.40$\pm$0.02 &  4.19$\pm$0.18 & 82.9 &   5.57$\pm$0.44 & $-$1.75$\pm$0.17 &  0.40$\pm$0.02 &  4.18$\pm$0.19 & 82.8 & 69.2\\
12&  6.75$\pm$0.92 & $-$2.12$\pm$0.31 &  0.33$\pm$0.15 &  3.91$\pm$0.32 & 76.2 &   6.75$\pm$0.93 & $-$2.13$\pm$0.31 &  0.33$\pm$0.15 &  3.91$\pm$0.32 & 75.9 & 71.9\\
12&  7.53$\pm$1.48 & $-$2.40$\pm$0.32 & $-$0.44$\pm$0.10 &  3.75$\pm$0.32 & 20.5 &   7.53$\pm$1.48 & $-$2.41$\pm$0.34 & $-$0.45$\pm$0.11 &  3.73$\pm$0.35 & 20.8 &      \\
13&  5.00$\pm$0.13 & $-$1.52$\pm$0.16 &  0.40          &  4.93$\pm$0.19 & 52.1 &   5.01$\pm$0.13 & $-$1.51$\pm$0.17 &  0.40          &  4.92$\pm$0.20 & 53.6 & 51.6\\
13&  6.36$\pm$0.17 & $-$1.68$\pm$0.07 & $-$0.40$\pm$0.01 &  4.93$\pm$0.07 & 47.4 &   6.35$\pm$0.18 & $-$1.68$\pm$0.07 & $-$0.40$\pm$0.01 &  4.93$\pm$0.07 & 46.0 &     \\
14&  7.01$\pm$1.11 & $-$1.70$\pm$0.47 &  0.28$\pm$0.18 &  4.20$\pm$0.50 & 60.6 &   7.00$\pm$1.11 & $-$1.71$\pm$0.47 &  0.29$\pm$0.18 &  4.19$\pm$0.50 & 60.7 & 77.3\\
14&  7.64$\pm$0.68 & $-$2.02$\pm$0.17 & $-$0.41$\pm$0.06 &  4.12$\pm$0.17 & 39.2 &   7.63$\pm$0.67 & $-$2.02$\pm$0.17 & $-$0.41$\pm$0.06 &  4.12$\pm$0.17 & 39.1 &     \\
15&  8.80$\pm$0.25 & $-$1.16$\pm$0.16 &  0.00          &  5.11$\pm$0.19 & 99.3 &   8.80$\pm$0.25 & $-$1.16$\pm$0.16 &  0.00          &  5.11$\pm$0.19 & 99.3 & 69.2\\
16& 10.40$\pm$0.47 & $-$0.81$\pm$0.15 &  0.00          &  5.63$\pm$0.19 & 89.1 &  10.40$\pm$0.47 & $-$0.82$\pm$0.15 &  0.00          &  5.62$\pm$0.19 & 89.1 & 56.0\\
18&  4.64$\pm$0.51 & $-$1.69$\pm$0.09 & $-$0.66$\pm$0.11 &  4.98$\pm$0.09 & 84.5 &   4.59$\pm$0.48 & $-$1.70$\pm$0.09 & $-$0.66$\pm$0.11 &  4.97$\pm$0.09 & 80.6 & 68.2\\
19& 10.17$\pm$0.37 & $-$1.08$\pm$0.10 &  0.00          &  5.47$\pm$0.12 & 99.8 &  10.17$\pm$0.37 & $-$1.09$\pm$0.10 &  0.00          &  5.47$\pm$0.12 & 99.8 & 54.6\\
21&  8.71$\pm$0.05 & $-$0.80$\pm$0.19 &  0.00          &  5.12$\pm$0.24 & 99.8 &   8.71$\pm$0.05 & $-$0.80$\pm$0.18 &  0.00          &  5.11$\pm$0.24 & 99.8 & 42.0\\
23&  7.88$\pm$0.58 & $-$1.56$\pm$0.19 & $-$0.43$\pm$0.09 &  4.26$\pm$0.19 & 82.5 &   7.87$\pm$0.58 & $-$1.56$\pm$0.19 & $-$0.43$\pm$0.09 &  4.26$\pm$0.19 & 81.8 & 69.5\\
26& 10.07$\pm$1.36 & $-$1.40$\pm$0.05 & $-$0.54$\pm$0.15 &  6.36$\pm$0.05 & 78.0 &  10.05$\pm$1.37 & $-$1.40$\pm$0.05 & $-$0.54$\pm$0.15 &  6.35$\pm$0.05 & 78.0 & 82.6\\
26&  7.61$\pm$0.41 & $-$1.25$\pm$0.19 &  0.00          &  6.33$\pm$0.23 & 21.5 &   7.61$\pm$0.41 & $-$1.26$\pm$0.19 &  0.00          &  6.33$\pm$0.23 & 21.5 &     \\
27&  4.66$\pm$0.63 & $-$2.16$\pm$0.09 & $-$0.48$\pm$0.30 &  5.03$\pm$0.09 & 64.0 &   4.54$\pm$0.64 & $-$2.18$\pm$0.09 & $-$0.46$\pm$0.32 &  5.00$\pm$0.09 & 62.8 & 52.8\\
27&  2.73$\pm$0.37 & $-$2.28$\pm$0.05 &  0.40          &  4.91$\pm$0.05 & 25.2 &   2.71$\pm$0.38 & $-$2.29$\pm$0.05 &  0.40          &  4.90$\pm$0.05 & 26.7 &     \\
28&  9.86$\pm$0.53 & $-$1.14$\pm$0.13 &  0.00          &  5.18$\pm$0.15 & 78.3 &   9.86$\pm$0.53 & $-$1.15$\pm$0.13 &  0.00          &  5.18$\pm$0.15 & 78.3 & 64.7\\
29& 14.50$\pm$1.47 & $-$1.07$\pm$0.13 &  0.00          &  5.46$\pm$0.15 & 99.8 &  14.48$\pm$1.46 & $-$1.07$\pm$0.13 &  0.00          &  5.46$\pm$0.14 & 99.8 & 65.2\\
30&  9.39$\pm$0.71 & $-$1.17$\pm$0.19 &  0.00          &  5.21$\pm$0.22 & 75.1 &   9.39$\pm$0.71 & $-$1.17$\pm$0.19 &  0.00          &  5.21$\pm$0.22 & 75.4 & 68.2\\
32&  7.64$\pm$0.78 & $-$1.52$\pm$0.56 & $-$0.25$\pm$0.20 &  4.78$\pm$0.65 & 98.5 &   7.63$\pm$0.79 & $-$1.53$\pm$0.57 & $-$0.25$\pm$0.20 &  4.78$\pm$0.66 & 98.5 & 58.9\\
33&  9.03$\pm$0.49 & $-$0.75$\pm$0.33 &  0.00          &  5.32$\pm$0.41 & 99.8 &   9.03$\pm$0.49 & $-$0.75$\pm$0.33 &  0.00          &  5.32$\pm$0.41 & 99.8 & 44.4\\
34&  3.82$\pm$1.73 & $-$1.61$\pm$1.06 &  0.37$\pm$0.10 &  3.48$\pm$1.26 & 46.3 &   3.81$\pm$1.72 & $-$1.61$\pm$1.05 &  0.37$\pm$0.10 &  3.48$\pm$1.25 & 45.7 & 76.7\\
34&  5.81$\pm$0.88 & $-$1.51$\pm$0.65 & $-$0.46$\pm$0.12 &  3.89$\pm$0.71 & 32.1 &   5.79$\pm$0.86 & $-$1.52$\pm$0.64 & $-$0.45$\pm$0.12 &  3.89$\pm$0.70 & 31.7 &     \\
34&  7.66$\pm$2.11 & $-$1.90$\pm$0.38 & $-$1.70          &  3.71$\pm$0.38 & 21.6 &   7.53$\pm$2.18 & $-$1.92$\pm$0.40 & $-$1.70          &  3.69$\pm$0.40 & 22.6 &     \\
37&  8.23$\pm$0.17 & $-$0.93$\pm$0.24 &  0.00          &  5.76$\pm$0.30 & 86.0 &   8.23$\pm$0.17 & $-$0.93$\pm$0.24 &  0.00          &  5.76$\pm$0.30 & 86.4 & 57.2\\
40&  7.60$\pm$1.10 & $-$1.28$\pm$0.20 & $-$0.24$\pm$0.32 &  5.63$\pm$0.24 & 95.5 &   6.84$\pm$0.63 & $-$1.29$\pm$0.24 &  0.01$\pm$0.07 &  5.60$\pm$0.28 & 62.6 & 80.0\\
40&                &                &                &                &      &   8.86$\pm$0.67 & $-$1.28$\pm$0.10 & $-$0.66$\pm$0.10 &  5.68$\pm$0.11 & 34.4 &     \\
42& 12.92$\pm$0.93 & $-$1.31$\pm$0.04 & $-$0.45$\pm$0.11 &  5.37$\pm$0.04 & 89.8 &  12.91$\pm$0.95 & $-$1.31$\pm$0.04 & $-$0.45$\pm$0.11 &  5.56$\pm$0.05 & 89.8 & 51.3\\
43&  9.06$\pm$0.30 & $-$0.73$\pm$0.25 &  0.00          &  5.16$\pm$0.32 & 99.8 &   9.05$\pm$0.29 & $-$0.73$\pm$0.25 &  0.00          &  5.18$\pm$0.32 & 99.8 & 48.3\\
45& 19.32$\pm$0.97 & $-$0.80$\pm$0.06 & $-$0.70          &  6.21$\pm$0.07 & 97.3 &  19.33$\pm$0.98 & $-$0.80$\pm$0.06 & $-$0.70          &  6.03$\pm$0.07 & 97.4 & 73.9\\
47&  5.13$\pm$0.17 & $-$1.92$\pm$0.11 &  0.40          &  4.43$\pm$0.12 & 93.2 &   5.14$\pm$0.18 & $-$1.92$\pm$0.11 &  0.40          &  4.60$\pm$0.12 & 93.8 & 50.2\\
48&  9.62$\pm$0.89 & $-$1.34$\pm$0.07 & $-$0.53$\pm$0.15 &  5.18$\pm$0.08 & 39.6 &   7.41$\pm$0.39 & $-$1.29$\pm$0.21 &  0.00          &  5.10$\pm$0.25 & 55.8 & 49.5\\
48&  7.41$\pm$0.39 & $-$1.29$\pm$0.21 &  0.00          &  5.11$\pm$0.25 & 55.4 &   9.60$\pm$0.89 & $-$1.34$\pm$0.08 & $-$0.53$\pm$0.15 &  5.19$\pm$0.08 & 39.3 &     \\
49& 13.46$\pm$0.67 & $-$1.34$\pm$0.14 & $-$1.70          &  4.74$\pm$0.14 & 35.2 &   8.25$\pm$1.26 & $-$1.34$\pm$0.29 & $-$0.36$\pm$0.27 &  4.48$\pm$0.34 & 64.6 & 59.6\\
49&  8.27$\pm$1.25 & $-$1.34$\pm$0.29 & $-$0.36$\pm$0.27 &  4.49$\pm$0.34 & 64.5 &  13.44$\pm$0.67 & $-$1.34$\pm$0.14 & $-$1.70          &  4.66$\pm$0.14 & 35.1 &     \\
52&  4.46$\pm$0.17 & $-$1.53$\pm$0.12 &  0.40          &  5.24$\pm$0.14 & 47.6 &   4.75$\pm$0.57 & $-$1.53$\pm$0.11 &  0.31$\pm$0.17 &  5.25$\pm$0.13 & 63.9 & 76.7\\
52& 11.89$\pm$0.47 & $-$1.19$\pm$0.03 & $-$1.70          &  5.82$\pm$0.03 & 32.5 &  11.88$\pm$0.49 & $-$1.19$\pm$0.03 & $-$1.70          &  5.71$\pm$0.03 & 31.1 & \\
53&  9.69$\pm$0.47 & $-$1.69$\pm$0.10 &  0.00          &  4.58$\pm$0.11 & 90.2 &   9.68$\pm$0.47 & $-$1.69$\pm$0.10 &  0.00          &  4.65$\pm$0.10 & 90.3 & 65.5\\
56&  9.25$\pm$0.90 & $-$1.61$\pm$0.18 &  0.04$\pm$0.13 &  4.18$\pm$0.19 & 62.7 &   9.24$\pm$0.90 & $-$1.61$\pm$0.18 &  0.04$\pm$0.13 &  4.19$\pm$0.19 & 62.6 & 63.4\\
56& 11.64$\pm$0.95 & $-$1.79$\pm$0.15 & $-$0.56$\pm$0.15 &  4.15$\pm$0.15 & 24.9 &  11.67$\pm$0.93 & $-$1.78$\pm$0.15 & $-$0.56$\pm$0.15 &  4.15$\pm$0.15 & 24.8 & \\
58&  4.86$\pm$0.32 & $-$1.76$\pm$0.13 &  0.37$\pm$0.11 &  4.61$\pm$0.15 & 76.1 &   4.86$\pm$0.32 & $-$1.75$\pm$0.13 &  0.37$\pm$0.10 &  4.57$\pm$0.15 & 75.5 & 70.4 \\
59& 10.42$\pm$0.39 & $-$1.32$\pm$0.09 &  0.00          &  4.95$\pm$0.10 & 60.0 &  10.42$\pm$0.39 & $-$1.32$\pm$0.09 &  0.00          &  4.91$\pm$0.10 & 60.2 & 65.5\\
59& 12.19$\pm$0.53 & $-$1.78$\pm$0.06 & $-$0.69$\pm$0.06 &  4.69$\pm$0.06 & 39.9 &  12.18$\pm$0.53 & $-$1.78$\pm$0.06 & $-$0.69$\pm$0.06 &  4.72$\pm$0.06 & 39.7 &     \\
60&  8.77$\pm$1.32 & $-$2.29$\pm$0.35 & $-$0.28$\pm$0.26 &  3.54$\pm$0.36 & 70.3 &   8.78$\pm$1.32 & $-$2.29$\pm$0.35 & $-$0.28$\pm$0.26 &  3.58$\pm$0.36 & 70.4 & 65.5\\
60&  6.47$\pm$0.71 & $-$2.43$\pm$0.33 &  0.40          &  3.41$\pm$0.33 & 25.3 &   6.49$\pm$0.63 & $-$2.42$\pm$0.32 &  0.40          &  3.38$\pm$0.32 & 25.3 &     \\
62&  8.86$\pm$0.56 & $-$1.27$\pm$0.24 &  0.00          &  4.73$\pm$0.27 & 75.7 &   8.85$\pm$0.55 & $-$1.27$\pm$0.23 &  0.00          &  4.71$\pm$0.26 & 75.9 & 51.1\\
64&  5.19$\pm$0.26 & $-$1.61$\pm$0.15 &  0.39$\pm$0.05 &  4.26$\pm$0.17 & 63.2 &   5.20$\pm$0.24 & $-$1.60$\pm$0.15 &  0.39$\pm$0.05 &  4.41$\pm$0.17 & 63.1 & 51.5\\
64&  6.05$\pm$0.30 & $-$1.81$\pm$0.15 & $-$0.41$\pm$0.06 &  4.40$\pm$0.15 & 32.8 &   6.02$\pm$0.32 & $-$1.82$\pm$0.15 & $-$0.42$\pm$0.07 &  4.21$\pm$0.16 & 32.5 &     \\
65& 12.87$\pm$1.91 & $-$2.19$\pm$0.32 & $-$1.70          &  3.77$\pm$0.32 & 48.5 &  12.80$\pm$2.08 & $-$2.20$\pm$0.35 & $-$1.70          &  3.76$\pm$0.35 & 48.4 & 50.2\\
65&  6.35$\pm$2.55 & $-$2.80$\pm$0.66 & $-$0.39$\pm$0.26 &  3.10$\pm$0.66 & 28.1 &   6.39$\pm$2.51 & $-$2.80$\pm$0.66 & $-$0.38$\pm$0.26 &  3.24$\pm$0.66 & 28.7 &     \\
65&  5.14$\pm$1.57 & $-$2.76$\pm$0.51 &  0.40          &  3.24$\pm$0.51 & 23.4 &   5.19$\pm$1.54 & $-$2.74$\pm$0.51 &  0.40          &  3.17$\pm$0.51 & 22.9 &     \\
66&  6.56$\pm$0.53 & $-$2.10$\pm$0.16 & $-$0.43$\pm$0.15 &  4.46$\pm$0.16 & 74.5 &   6.55$\pm$0.54 & $-$2.10$\pm$0.16 & $-$0.43$\pm$0.15 &  4.55$\pm$0.16 & 74.4 & 78.5\\
68& 10.81$\pm$1.25 & $-$1.87$\pm$0.20 & $-$0.27$\pm$0.34 &  4.18$\pm$0.21 & 99.5 &  10.80$\pm$1.24 & $-$1.87$\pm$0.20 & $-$0.27$\pm$0.34 &  4.29$\pm$0.21 & 99.5 & 58.9\\
70& 12.00$\pm$0.48 & $-$1.02$\pm$0.04 & $-$1.70          &  5.94$\pm$0.04 & 67.7 &  11.99$\pm$0.48 & $-$1.03$\pm$0.04 & $-$1.70          &  5.72$\pm$0.04 & 67.0 & 78.0\\
70&  5.12$\pm$0.76 & $-$1.37$\pm$0.12 &  0.18$\pm$0.20 &  5.26$\pm$0.13 & 27.9 &   5.06$\pm$0.76 & $-$1.37$\pm$0.12 &  0.19$\pm$0.20 &  5.18$\pm$0.13 & 28.6 &     \\
74& 13.13$\pm$0.67 & $-$1.62$\pm$0.11 & $-$1.70          &  4.56$\pm$0.11 & 63.6 &  13.10$\pm$0.67 & $-$1.62$\pm$0.11 & $-$1.70          &  4.76$\pm$0.11 & 63.9 & 72.3\\
74&  7.71$\pm$1.16 & $-$1.73$\pm$0.26 & $-$0.25$\pm$0.29 &  4.55$\pm$0.29 & 35.3 &   7.68$\pm$1.17 & $-$1.73$\pm$0.26 & $-$0.24$\pm$0.29 &  4.45$\pm$0.29 & 34.9 &     \\
75&  9.50$\pm$1.45 & $-$1.73$\pm$0.25 & $-$0.36$\pm$0.29 &  4.17$\pm$0.27 & 71.5 &   9.51$\pm$1.45 & $-$1.73$\pm$0.25 & $-$0.36$\pm$0.29 &  4.17$\pm$0.27 & 71.3 & 68.8\\
75& 14.82$\pm$1.04 & $-$1.68$\pm$0.16 & $-$1.70          &  4.31$\pm$0.16 & 27.3 &  14.81$\pm$1.04 & $-$1.68$\pm$0.16 & $-$1.70          &  4.29$\pm$0.16 & 27.5 &     \\
76&  7.25$\pm$1.21 & $-$1.92$\pm$0.39 &  0.17$\pm$0.20 &  3.85$\pm$0.40 & 47.2 &   7.28$\pm$1.18 & $-$1.91$\pm$0.38 &  0.17$\pm$0.20 &  3.93$\pm$0.39 & 46.2 & 48.0\\
76&  8.63$\pm$1.19 & $-$2.09$\pm$0.22 & $-$0.50$\pm$0.14 &  3.87$\pm$0.22 & 29.0 &   8.57$\pm$1.24 & $-$2.09$\pm$0.23 & $-$0.50$\pm$0.14 &  3.99$\pm$0.23 & 29.1 &     \\
76& 13.03$\pm$1.17 & $-$1.96$\pm$0.20 & $-$1.70          &  4.15$\pm$0.20 & 23.8 &  13.04$\pm$1.00 & $-$1.96$\pm$0.17 & $-$1.70          &  3.91$\pm$0.17 & 24.7 &     \\
77&  6.74$\pm$1.26 & $-$2.14$\pm$0.42 & $-$0.43$\pm$0.16 &  2.99$\pm$0.43 & 64.8 &   6.73$\pm$1.29 & $-$2.14$\pm$0.43 & $-$0.43$\pm$0.17 &  3.12$\pm$0.43 & 64.3 & 66.0\\
78&  5.79$\pm$1.81 & $-$2.59$\pm$0.51 & $-$0.45$\pm$0.24 &  2.85$\pm$0.51 & 51.1 &   6.02$\pm$1.90 & $-$2.53$\pm$0.51 & $-$0.52$\pm$0.15 &  2.98$\pm$0.51 & 41.8 & 46.3\\
78&  8.25$\pm$4.88 & $-$2.50$\pm$0.78 & $-$1.70          &  2.81$\pm$0.78 & 33.1 &   8.10$\pm$4.84 & $-$2.52$\pm$0.77 & $-$1.70          &  2.78$\pm$0.77 & 33.8 &     \\
78&                &                &                &                &      &   3.91$\pm$1.38 & $-$2.79$\pm$0.39 &  0.27$\pm$0.19 &  2.69$\pm$0.39 & 24.4 &     \\
80&  6.18$\pm$0.46 & $-$1.67$\pm$0.26 & $-$0.50$\pm$0.14 &  4.28$\pm$0.27 & 61.0 &   6.17$\pm$0.46 & $-$1.67$\pm$0.26 & $-$0.50$\pm$0.14 &  4.27$\pm$0.27 & 59.7 & 72.6\\
80&  4.27$\pm$0.81 & $-$1.96$\pm$0.23 &  0.26$\pm$0.19 &  4.01$\pm$0.24 & 22.1 &   4.21$\pm$0.80 & $-$1.97$\pm$0.23 &  0.26$\pm$0.19 &  3.84$\pm$0.25 & 24.5 &     \\
\hline
\end{tabular}
\end{bf}
\end{tiny}
\label{tableres}
\end{table*}

\begin{table*}
\addtocounter{table}{-1}
\caption[]{(cont.) Mean value and standard deviation for the age, 
burst strength, mass and metallicity of each individual cluster of
solutions. Only the properties of those clusters with
probability higher than 20~per cent for each region in column~1 are
given. Results for models with 15~per cent and a null fraction of Ly
photons escaping from the galaxy are shown separately. Probabilities
for these clusters of solutions are given in columns 6 and 11.
Metallicity is expressed as log($Z$/$Z_{\sun}$).}
\begin{tiny}
\begin{bf}
\begin{tabular}{r|ccccc|ccccc}
\hline
\multicolumn{11}{c}{Sub-solar metallicity}\\
\hline
& \multicolumn{5}{|c|}{15\% escaping Ly photons} & \multicolumn{5}{|c}{0\% escaping Ly photons} \\
 \# & Age & log $b$ & $Z$ &   log $M$  & Probab. & Age & log $b$ & $Z$ &   log $M$  & Probab. \\
    &(Myr)&         &     &(M$_{\sun}$)&    (\%)     &(Myr)&         &     &(M$_{\sun}$)&    (\%)     \\
\hline
6 &  6.32$\pm$0.29 & $-$1.88$\pm$0.19 & $-$0.40 & 3.88$\pm$0.19 &78.9&    6.31$\pm$0.29 & $-$1.89$\pm$0.19 & $-$0.40 & 3.87$\pm$0.19 & 78.0\\
7 &  6.95$\pm$0.62 & $-$2.48$\pm$0.19 & $-$0.40 & 3.71$\pm$0.19 &80.7&    6.92$\pm$0.68 & $-$2.49$\pm$0.20 & $-$0.40 & 3.70$\pm$0.20 & 81.1\\
8 &  8.91$\pm$0.79 & $-$1.99$\pm$0.14 & $-$1.70 & 4.19$\pm$0.14 &55.2&    8.84$\pm$0.79 & $-$2.00$\pm$0.14 & $-$1.70 & 4.21$\pm$0.14 & 57.8\\
8 &  5.90$\pm$0.45 & $-$2.11$\pm$0.16 & $-$0.40 & 4.04$\pm$0.16 &39.5&    5.93$\pm$0.41 & $-$2.10$\pm$0.15 & $-$0.40 & 4.05$\pm$0.15 & 36.8\\
12&  7.34$\pm$0.95 & $-$2.43$\pm$0.28 & $-$0.40 & 3.71$\pm$0.28 &44.5&    7.21$\pm$1.10 & $-$2.47$\pm$0.32 & $-$0.40 & 3.68$\pm$0.32 & 44.9\\
12& 10.66$\pm$2.40 & $-$2.45$\pm$0.40 & $-$1.70 & 3.71$\pm$0.40 &31.4&   10.88$\pm$2.25 & $-$2.42$\pm$0.37 & $-$1.70 & 3.74$\pm$0.37 & 30.3\\
12&  9.48$\pm$1.57 & $-$2.17$\pm$0.33 & $-$0.70 & 3.97$\pm$0.33 &24.1&    9.45$\pm$1.49 & $-$2.19$\pm$0.34 & $-$0.70 & 4.10$\pm$0.34 & 24.8\\
13&  6.31$\pm$0.18 & $-$1.70$\pm$0.08 & $-$0.40 & 4.92$\pm$0.08 &96.6&    6.30$\pm$0.19 & $-$1.70$\pm$0.08 & $-$0.40 & 4.92$\pm$0.08 & 96.6\\
14&  7.47$\pm$0.61 & $-$2.07$\pm$0.18 & $-$0.40 & 4.07$\pm$0.18 &81.4&    7.46$\pm$0.61 & $-$2.07$\pm$0.18 & $-$0.40 & 4.07$\pm$0.18 & 81.6\\
15& 10.19$\pm$0.31 & $-$1.82$\pm$0.06 & $-$0.70 & 4.81$\pm$0.06 &59.4&   10.21$\pm$0.33 & $-$1.82$\pm$0.06 & $-$0.70 & 4.76$\pm$0.06 & 58.1\\
15&  8.33$\pm$0.60 & $-$1.95$\pm$0.09 & $-$0.40 & 4.68$\pm$0.09 &40.5&    8.27$\pm$0.56 & $-$1.96$\pm$0.09 & $-$0.40 & 4.67$\pm$0.09 & 41.8\\
16& 11.81$\pm$0.42 & $-$1.60$\pm$0.04 & $-$0.70 & 5.35$\pm$0.04 &80.9&   11.79$\pm$0.43 & $-$1.60$\pm$0.04 & $-$0.70 & 5.35$\pm$0.04 & 80.9\\
18&  4.53$\pm$0.43 & $-$1.71$\pm$0.08 & $-$0.70 & 4.96$\pm$0.08 &74.0&    4.49$\pm$0.39 & $-$1.72$\pm$0.08 & $-$0.70 & 4.94$\pm$0.08 & 70.0\\
19& 11.76$\pm$0.41 & $-$1.72$\pm$0.04 & $-$0.70 & 5.18$\pm$0.04 &90.0&   11.77$\pm$0.42 & $-$1.73$\pm$0.04 & $-$0.70 & 5.18$\pm$0.04 & 91.4\\
21&  7.60$\pm$0.29 & $-$1.96$\pm$0.05 & $-$0.40 & 4.56$\pm$0.05 &72.3&    7.63$\pm$0.30 & $-$1.96$\pm$0.05 & $-$0.40 & 4.56$\pm$0.05 & 77.2\\
21& 10.00$\pm$0.09 & $-$1.79$\pm$0.04 & $-$0.70 & 4.74$\pm$0.04 &27.6&    9.98$\pm$0.18 & $-$1.79$\pm$0.04 & $-$0.70 & 4.51$\pm$0.04 & 22.7 \\
23&  7.79$\pm$0.46 & $-$1.55$\pm$0.18 & $-$0.40 & 4.26$\pm$0.18 &73.8&    7.77$\pm$0.46 & $-$1.55$\pm$0.18 & $-$0.40 & 4.18$\pm$0.18 & 73.8\\
26&  8.96$\pm$0.70 & $-$1.42$\pm$0.04 & $-$0.40 & 6.32$\pm$0.05 &52.5&    8.94$\pm$0.70 & $-$1.43$\pm$0.05 & $-$0.40 & 6.52$\pm$0.05 & 52.2\\
26& 11.17$\pm$0.59 & $-$1.36$\pm$0.03 & $-$0.70 & 6.40$\pm$0.03 &46.8&   11.15$\pm$0.63 & $-$1.36$\pm$0.04 & $-$0.70 & 6.40$\pm$0.04 & 47.2\\
27&  4.91$\pm$0.58 & $-$2.15$\pm$0.11 & $-$0.70 & 5.03$\pm$0.11 &44.2&    7.79$\pm$0.78 & $-$1.82$\pm$0.05 & $-$1.70 & 5.28$\pm$0.05 & 45.4\\
27&  7.90$\pm$0.70 & $-$1.81$\pm$0.05 & $-$1.70 & 5.39$\pm$0.05 &43.6&    4.84$\pm$0.59 & $-$2.17$\pm$0.11 & $-$0.70 & 5.01$\pm$0.11 & 44.4\\
28& 11.44$\pm$0.38 & $-$1.74$\pm$0.04 & $-$0.70 & 4.91$\pm$0.04 &88.7&   11.44$\pm$0.38 & $-$1.74$\pm$0.04 & $-$0.70 & 4.91$\pm$0.04 & 88.3\\
29& 18.66$\pm$0.97 & $-$1.42$\pm$0.05 & $-$0.70 & 5.35$\pm$0.05 &85.1&   18.67$\pm$0.98 & $-$1.42$\pm$0.05 & $-$0.70 & 5.33$\pm$0.05 & 84.6\\
30& 11.24$\pm$0.32 & $-$1.73$\pm$0.04 & $-$0.70 & 4.96$\pm$0.04 &59.5&   11.23$\pm$0.33 & $-$1.73$\pm$0.04 & $-$0.70 & 5.15$\pm$0.04 & 60.0\\
30&  8.84$\pm$0.32 & $-$1.84$\pm$0.05 & $-$0.40 & 4.86$\pm$0.05 &40.4&    8.84$\pm$0.32 & $-$1.84$\pm$0.05 & $-$0.40 & 4.92$\pm$0.05 & 39.9\\
32&  7.15$\pm$0.35 & $-$1.93$\pm$0.05 & $-$0.40 & 4.60$\pm$0.05 &97.8&    7.14$\pm$0.35 & $-$1.93$\pm$0.05 & $-$0.40 & 4.48$\pm$0.05 & 97.9\\
33& 10.43$\pm$0.37 & $-$1.66$\pm$0.04 & $-$0.70 & 5.02$\pm$0.04 &70.7&   10.44$\pm$0.36 & $-$1.65$\pm$0.04 & $-$0.70 & 4.79$\pm$0.05 & 73.5\\
33&  8.97$\pm$0.51 & $-$1.73$\pm$0.05 & $-$0.40 & 4.95$\pm$0.05 &29.2&    8.95$\pm$0.56 & $-$1.73$\pm$0.05 & $-$0.40 & 4.71$\pm$0.06 & 26.4\\
34&  7.97$\pm$1.79 & $-$1.85$\pm$0.34 & $-$1.70 & 3.76$\pm$0.34 &54.0&    7.90$\pm$1.82 & $-$1.86$\pm$0.35 & $-$1.70 & 3.88$\pm$0.35 & 55.6\\
34&  5.75$\pm$0.93 & $-$1.50$\pm$0.68 & $-$0.40 & 3.90$\pm$0.74 &37.7&    5.75$\pm$0.93 & $-$1.50$\pm$0.66 & $-$0.40 & 3.88$\pm$0.71 & 36.6\\
37&  7.79$\pm$0.43 & $-$1.68$\pm$0.05 & $-$0.40 & 5.47$\pm$0.05 &75.7&    7.77$\pm$0.43 & $-$1.69$\pm$0.05 & $-$0.40 & 5.32$\pm$0.05 & 74.2\\
37& 10.02$\pm$0.22 & $-$1.54$\pm$0.04 & $-$0.70 & 5.61$\pm$0.04 &24.2&   10.02$\pm$0.21 & $-$1.54$\pm$0.04 & $-$0.70 & 5.66$\pm$0.04 & 25.7\\
40&  9.13$\pm$0.78 & $-$1.26$\pm$0.12 & $-$0.70 & 5.70$\pm$0.13 &73.3&    9.13$\pm$0.78 & $-$1.26$\pm$0.12 & $-$0.70 & 5.92$\pm$0.13 & 74.0\\
42& 12.67$\pm$0.67 & $-$1.31$\pm$0.04 & $-$0.40 & 5.55$\pm$0.04 &78.0&   12.66$\pm$0.67 & $-$1.31$\pm$0.04 & $-$0.40 & 5.39$\pm$0.04 & 77.9\\
43& 10.46$\pm$0.37 & $-$1.69$\pm$0.06 & $-$0.70 & 4.85$\pm$0.06 &68.5&   10.49$\pm$0.35 & $-$1.69$\pm$0.05 & $-$0.70 & 4.84$\pm$0.05 & 73.8\\
43&  9.39$\pm$0.51 & $-$1.73$\pm$0.06 & $-$0.40 & 4.82$\pm$0.07 &31.4&    9.43$\pm$0.56 & $-$1.73$\pm$0.07 & $-$0.40 & 4.81$\pm$0.07 & 26.1\\
45& 19.32$\pm$0.97 & $-$0.80$\pm$0.06 & $-$0.70 & 6.03$\pm$0.07 &97.9&   19.33$\pm$0.98 & $-$0.80$\pm$0.06 & $-$0.70 & 6.20$\pm$0.07 & 98.0\\
47& 10.72$\pm$0.52 & $-$1.79$\pm$0.06 & $-$1.70 & 4.86$\pm$0.06 &47.0&   10.69$\pm$0.53 & $-$1.79$\pm$0.06 & $-$1.70 & 4.78$\pm$0.06 & 47.9\\
47&  6.40$\pm$0.27 & $-$2.02$\pm$0.09 & $-$0.40 & 4.59$\pm$0.09 &45.2&    6.39$\pm$0.27 & $-$2.03$\pm$0.09 & $-$0.40 & 4.51$\pm$0.09 & 45.5\\
48& 10.32$\pm$0.49 & $-$1.30$\pm$0.07 & $-$0.70 & 5.22$\pm$0.07 &53.0&   10.31$\pm$0.49 & $-$1.30$\pm$0.07 & $-$0.70 & 5.02$\pm$0.07 & 53.0\\
48&  9.30$\pm$0.64 & $-$1.34$\pm$0.07 & $-$0.40 & 5.18$\pm$0.07 &36.6&    9.29$\pm$0.65 & $-$1.34$\pm$0.07 & $-$0.40 & 4.98$\pm$0.07 & 36.7\\
49& 13.46$\pm$0.65 & $-$1.34$\pm$0.13 & $-$1.70 & 4.66$\pm$0.13 &44.3&   13.44$\pm$0.65 & $-$1.34$\pm$0.14 & $-$1.70 & 4.62$\pm$0.14 & 44.2\\
49&  9.81$\pm$0.65 & $-$1.28$\pm$0.20 & $-$0.70 & 4.57$\pm$0.21 &27.9&    9.80$\pm$0.65 & $-$1.28$\pm$0.20 & $-$0.70 & 4.54$\pm$0.21 & 28.2\\
49&  8.39$\pm$0.51 & $-$1.32$\pm$0.18 & $-$0.40 & 4.52$\pm$0.19 &27.8&    8.38$\pm$0.52 & $-$1.31$\pm$0.18 & $-$0.40 & 4.49$\pm$0.19 & 27.6\\
52& 11.60$\pm$0.52 & $-$1.20$\pm$0.03 & $-$1.70 & 5.72$\pm$0.03 &75.2&   11.57$\pm$0.52 & $-$1.20$\pm$0.03 & $-$1.70 & 5.78$\pm$0.03 & 74.7\\
52&  7.47$\pm$0.46 & $-$1.42$\pm$0.07 & $-$0.70 & 5.41$\pm$0.07 &24.7&    7.45$\pm$0.46 & $-$1.43$\pm$0.07 & $-$0.70 & 5.44$\pm$0.07 & 25.2\\
53& 11.46$\pm$0.47 & $-$2.05$\pm$0.08 & $-$0.70 & 4.41$\pm$0.08 &66.8&   11.47$\pm$0.46 & $-$2.05$\pm$0.08 & $-$0.70 & 4.42$\pm$0.08 & 66.8\\
53& 10.21$\pm$0.82 & $-$2.08$\pm$0.12 & $-$0.40 & 4.38$\pm$0.12 &33.1&   10.17$\pm$0.84 & $-$2.09$\pm$0.12 & $-$0.40 & 4.34$\pm$0.12 & 33.1\\
56& 10.96$\pm$0.65 & $-$1.81$\pm$0.14 & $-$0.40 & 4.13$\pm$0.14 &66.3&   10.96$\pm$0.66 & $-$1.81$\pm$0.14 & $-$0.40 & 4.09$\pm$0.14 & 66.0\\
56& 12.01$\pm$0.80 & $-$1.83$\pm$0.16 & $-$0.70 & 4.12$\pm$0.16 &20.6&   12.02$\pm$0.80 & $-$1.82$\pm$0.16 & $-$0.70 & 4.07$\pm$0.16 & 20.7\\
58& 10.72$\pm$0.54 & $-$1.50$\pm$0.07 & $-$1.70 & 4.97$\pm$0.07 &63.5&   10.69$\pm$0.56 & $-$1.50$\pm$0.07 & $-$1.70 & 5.10$\pm$0.07 & 64.1 \\
59& 12.19$\pm$0.49 & $-$1.79$\pm$0.06 & $-$0.70 & 4.71$\pm$0.06 &82.7&   12.19$\pm$0.49 & $-$1.79$\pm$0.06 & $-$0.70 & 4.64$\pm$0.06 & 82.8\\
60&  8.36$\pm$1.24 & $-$2.51$\pm$0.31 & $-$0.40 & 3.38$\pm$0.31 &53.6&    8.30$\pm$1.24 & $-$2.52$\pm$0.31 & $-$0.40 & 3.44$\pm$0.31 & 53.7\\
60& 10.42$\pm$1.35 & $-$2.31$\pm$0.30 & $-$0.70 & 3.57$\pm$0.30 &34.2&   10.46$\pm$1.29 & $-$2.30$\pm$0.28 & $-$0.70 & 3.56$\pm$0.28 & 34.4\\
62&  8.98$\pm$0.47 & $-$1.84$\pm$0.09 & $-$0.40 & 4.57$\pm$0.09 &51.2&    8.96$\pm$0.48 & $-$1.85$\pm$0.09 & $-$0.40 & 4.44$\pm$0.09 & 50.5\\
62& 10.88$\pm$0.45 & $-$1.73$\pm$0.07 & $-$0.70 & 4.68$\pm$0.07 &48.7&   10.87$\pm$0.44 & $-$1.73$\pm$0.07 & $-$0.70 & 4.65$\pm$0.07 & 49.4\\
64&  6.06$\pm$0.27 & $-$1.80$\pm$0.13 & $-$0.40 & 4.40$\pm$0.14 &62.7&    6.04$\pm$0.28 & $-$1.81$\pm$0.13 & $-$0.40 & 4.49$\pm$0.14 & 61.9\\
64&  9.84$\pm$0.63 & $-$1.63$\pm$0.10 & $-$1.70 & 4.64$\pm$0.10 &32.4&    9.77$\pm$0.64 & $-$1.64$\pm$0.10 & $-$1.70 & 4.50$\pm$0.10 & 32.9\\
65& 12.35$\pm$2.05 & $-$2.28$\pm$0.35 & $-$1.70 & 3.82$\pm$0.35 &71.0&   12.29$\pm$2.16 & $-$2.29$\pm$0.37 & $-$1.70 & 3.59$\pm$0.37 & 70.4\\
66&  6.53$\pm$0.34 & $-$2.07$\pm$0.14 & $-$0.40 & 4.41$\pm$0.14 &61.4&    6.52$\pm$0.34 & $-$2.08$\pm$0.14 & $-$0.40 & 4.59$\pm$0.14 & 61.3\\
66& 10.82$\pm$0.80 & $-$1.88$\pm$0.12 & $-$1.70 & 4.65$\pm$0.12 &21.2&   10.82$\pm$0.77 & $-$1.88$\pm$0.12 & $-$1.70 & 4.73$\pm$0.12 & 21.6\\
68& 12.11$\pm$0.67 & $-$2.04$\pm$0.11 & $-$0.70 & 4.04$\pm$0.11 &74.6&   12.09$\pm$0.67 & $-$2.04$\pm$0.11 & $-$0.70 & 4.00$\pm$0.11 & 74.7\\
68& 10.92$\pm$1.11 & $-$2.06$\pm$0.19 & $-$0.40 & 4.02$\pm$0.19 &25.1&   10.93$\pm$1.10 & $-$2.05$\pm$0.18 & $-$0.40 & 3.96$\pm$0.18 & 25.0\\
70& 11.92$\pm$0.54 & $-$1.03$\pm$0.04 & $-$1.70 & 5.67$\pm$0.04 &83.9&   11.90$\pm$0.54 & $-$1.03$\pm$0.04 & $-$1.70 & 5.79$\pm$0.04 & 83.9\\
74& 13.16$\pm$0.67 & $-$1.61$\pm$0.11 & $-$1.70 & 4.62$\pm$0.11 &76.6&   13.13$\pm$0.68 & $-$1.62$\pm$0.11 & $-$1.70 & 4.83$\pm$0.11 & 77.1\\
75&  9.36$\pm$0.87 & $-$1.83$\pm$0.18 & $-$0.40 & 4.09$\pm$0.18 &37.6&    9.36$\pm$0.88 & $-$1.83$\pm$0.18 & $-$0.40 & 4.17$\pm$0.18 & 36.8\\
75& 11.08$\pm$0.66 & $-$1.71$\pm$0.15 & $-$0.70 & 4.21$\pm$0.15 &31.6&   11.09$\pm$0.62 & $-$1.71$\pm$0.14 & $-$0.70 & 4.16$\pm$0.14 & 31.7\\
75& 14.71$\pm$1.06 & $-$1.69$\pm$0.16 & $-$1.70 & 4.27$\pm$0.16 &30.8&   14.68$\pm$1.07 & $-$1.70$\pm$0.16 & $-$1.70 & 4.20$\pm$0.16 & 31.5\\
76& 12.80$\pm$1.23 & $-$2.00$\pm$0.22 & $-$1.70 & 4.08$\pm$0.22 &43.6&   12.84$\pm$1.10 & $-$2.00$\pm$0.19 & $-$1.70 & 4.08$\pm$0.19 & 43.7\\
76&  8.36$\pm$0.98 & $-$2.08$\pm$0.24 & $-$0.40 & 3.97$\pm$0.24 &29.8&    8.31$\pm$1.07 & $-$2.09$\pm$0.27 & $-$0.40 & 3.77$\pm$0.27 & 30.2\\
76&  9.88$\pm$0.90 & $-$1.97$\pm$0.19 & $-$0.70 & 4.08$\pm$0.19 &26.6&    9.88$\pm$0.90 & $-$1.97$\pm$0.19 & $-$0.70 & 4.05$\pm$0.19 & 26.1\\
77&  6.93$\pm$0.94 & $-$2.05$\pm$0.33 & $-$0.40 & 3.15$\pm$0.33 &56.2&    6.96$\pm$0.84 & $-$2.04$\pm$0.32 & $-$0.40 & 3.28$\pm$0.32 & 54.8\\
77&  9.97$\pm$3.26 & $-$2.13$\pm$0.54 & $-$1.70 & 3.12$\pm$0.54 &28.8&    9.92$\pm$3.25 & $-$2.14$\pm$0.54 & $-$1.70 & 3.23$\pm$0.54 & 29.8\\
78&  8.67$\pm$4.35 & $-$2.44$\pm$0.69 & $-$1.70 & 3.10$\pm$0.69 &45.6&    8.52$\pm$4.32 & $-$2.47$\pm$0.69 & $-$1.70 & 2.97$\pm$0.69 & 47.1\\
78&  6.35$\pm$1.13 & $-$2.40$\pm$0.39 & $-$0.40 & 3.12$\pm$0.39 &30.2&    6.28$\pm$1.37 & $-$2.41$\pm$0.42 & $-$0.40 & 2.97$\pm$0.42 & 30.3\\
78&  5.48$\pm$2.17 & $-$2.81$\pm$0.52 & $-$0.70 & 2.72$\pm$0.52 &24.2&    5.62$\pm$2.24 & $-$2.76$\pm$0.52 & $-$0.70 & 2.55$\pm$0.52 & 22.6\\
80&  6.09$\pm$0.31 & $-$1.61$\pm$0.25 & $-$0.40 & 4.13$\pm$0.27 &41.7&    6.08$\pm$0.29 & $-$1.61$\pm$0.25 & $-$0.40 & 4.18$\pm$0.26 & 41.1\\
80&  9.85$\pm$0.86 & $-$1.54$\pm$0.17 & $-$1.70 & 4.32$\pm$0.17 &31.9&    9.81$\pm$0.88 & $-$1.55$\pm$0.17 & $-$1.70 & 4.37$\pm$0.17 & 32.9\\
80&  6.38$\pm$0.62 & $-$1.80$\pm$0.21 & $-$0.70 & 4.03$\pm$0.21 &26.4&    6.34$\pm$0.63 & $-$1.81$\pm$0.22 & $-$0.70 & 4.04$\pm$0.22 & 26.0\\
\hline
\end{tabular}
\end{bf}
\end{tiny}
\end{table*}

Then, using these expressions and the index values for the underlying
and starburst populations --from the predictions of the SSP Bruzual \&
Charlot (priv. comm.) models--, we obtained the results shown in
Table~\ref{indexestable}. The indexes measured (column 5 in
Table~\ref{indexestable}) were corrected in order to take into account
the different spectral resolution between our spectra and those where
the Lick indexes were originally defined (see Gorgas et
al. \cite{gorgas93} and references therein) and also the fact that our
spectra are flux-calibrated.

Thus, the Mg2 index in the Lick system should be 0.02 magnitudes
higher than that measured on flux-calibrated spectra (J. Gorgas,
priv. comm.). On the other hand, small differences in the spectral
resolution relative to the Lick library spectra yield significant
changes in the atomic index values. The Lick library spectra show
resolutions ($\sigma_{\mathrm{instrumental}}$) of about
200\,km\,s$^{-1}$ for the Fe5270 and Fe5406 indexes (J. Gorgas,
priv. comm.), being $\sigma_{\mathrm{instrumental}}$ for our spectra
116\,km\,s$^{-1}$. The corrected indexes are shown in column 6 of
Table~\ref{indexestable}.

Despite of the results shown in this table are compatible with those
obtained from the optical-near-infrared colors analysis, the burst
strength derived seems to be slightly lower. This difference is
probably due to the difference in size between the region covered by
the slit \#4b and the aperture used to measure the starburst colors.

\begin{figure}
%\resizebox{\hsize}{!}{\includegraphics{fig3.eps}} 
\vspace{18.4cm}
\caption{Age, burst strength and stellar mass of the star-forming regions using a 
symbol size code overplotted on the H$\alpha$ image.}
\label{properties}
\end{figure}

Therefore, we can conclude that our data, both colors and
spectroscopic indexes, well agree with a scenario constituted by
a 30\,Myr old burst superimposed on a several Gyr old stellar
population. 

Using the mass-to-light ratio predicted for the composite stellar
population, the burst strength and the absolute magnitude measured
within the aperture we obtained the stellar mass for the
starburst. This mass was corrected using the $f$ ratio between the
knot continuum emission in the aperture and its total continuum
emission. The total mass derived for this component was
9$\times$10$^{6}$\,M$_{\sun}$ with an $f$ factor of 0.619 (61.9~per
cent). As we commented in Sect.~\ref{under} for the underlying
population, the age, burst strenght and color excess deduced for the
starburst component are not affected by the uncertainty in the
distance to \object{Mrk~86}. However, due to this distance
uncertainty, its stellar mass is not known with a precision better
than 40~per cent (for a 20~per cent distance uncertainty;
M. E. Sharina, priv. comm.).

\begin{table*}
\caption[]{Ionized gas diagnostic}
\begin{tabular}{c|rc|ccccccc}
\hline
\# & $n_{\mathrm{e}}$ & $T_{\mathrm{e}}$ & [O+/H] & [O++/H]4959 &[O++/H]5007 & [O++/H] & [O/H] & 12+log\,[O/H] & Z/Z$_{\sun}$\\  
     & (cm$^{-3}$) & (K) & & & & & & & \\
\hline
 26  & 100  & 10000 & 4.061($-$5)  & 1.642($-$5) &    1.713($-$5)  &  1.677($-$5) & 5.74($-$5)  & 7.76  &  1/11 \\
 32  &  53  & 16300 & 5.570($-$5)  & 1.737($-$5) &    1.822($-$5)  &  1.780($-$5) & 7.35($-$5)  & 7.87  &  1/9  \\
 42  & 100  & 14900 & 1.737($-$5)  & $\geq$6.236($-$6) & $\geq$6.185($-$6) &$\geq$6.210($-$6) &$\geq$2.36($-$5)&$\geq$7.37  & $\geq$1/28\\
 47  &  68  & 24250 &    --     & 4.668($-$5) &    5.274($-$5)  &  4.971($-$5) &  --      &  --   &   --  \\
 52  &  10  & 10000 & 3.001($-$5)  & 1.009($-$5) &    1.003($-$5)  &  1.006($-$5) & 4.01($-$5)  & 7.60  &  1/16 \\
 54  & 100  & 17600 &    --     &   --     &    5.199($-$5)  &  5.199($-$5) &  --      &  --   &   --  \\
 66  & 200  & 10000 & 4.460($-$5)  & 2.003($-$5) &    1.896($-$5)  &  1.950($-$5) & 6.41($-$5)  & 7.81  &  1/10 \\
 70  &  10  & 16700 & 1.752($-$5)  & 1.412($-$5) &    1.659($-$5)  &  1.536($-$5) & 3.29($-$5)  & 7.52  &  1/20 \\
     &  55  & 20300 &    --     &    --    &       --     &     --    &   --     &  --   &   -- \\    
\hline
     & $n_{\mathrm{e}}$ & $T_{\mathrm{e}}$ & [N+/H] & [O/N]          & y+5876        & y+6678 & y+     &   y      &   Y  \\
     &   (cm$^{-3}$)    &       (K)        &        &                &               &         &       &               &  \\
\hline
 26 &   53  & 16300 & 8.724($-$6) &  6.6  &   0.121  &   0.089 &0.105 & 0.105 &0.296 \\
 32 &  100  & 14900 &   --     &  --   &    --    &    --   & --   &  --   & --   \\
 42 &   68  & 24250 & 1.816($-$6) & 13.0  &   0.120  &    --   &0.120 & 0.120 &0.324 \\
 47 &   10  & 10000 & 1.222($-$5) &  --   &    --    &    --   & --   &  --   & --   \\
 52 &  100  & 17600 &   --     &  --   &    --    &    --   & --   &  --   & --   \\
 54 &  200  & 10000 & 1.861($-$5) &  --   &    --    &    --   & --   &  --   & --   \\
 66 &   10  & 16700 & 2.978($-$6) & 21.5  &   0.100  &   0.115 &0.108 & 0.108 &0.301 \\
 70 &   55  & 20300 & 2.131($-$6) & 15.4  &   0.148  &    --   &0.148 & 0.148 &0.372 \\
\hline
\end{tabular}
\label{zs}
\end{table*}

\subsection{Star-forming regions}
\label{starforming}

Using the maximization procedure described in
Sect.~\ref{formingmethod} we derived ages, burst strengths and stellar
masses for those regions showing H$\alpha$ emission (see Table~4 in
Paper~I). The $B-V$, $V-R$, $V-J$, $J-H$ and $J-K$ colors and
H$\alpha$ fluxes were compared with evolutionary synthesis
models. In this study, models {\bf 1)} with metallicity in the
range 1/50\,Z$_{\sun}$$<$Z$<$2\,Z$_{\sun}$ and {\bf 2)} lower than
solar, and {\bf 3)} with 15~per cent and {\bf 4)} a null fraction of
escaping Lyman photons were explored.

In Table~\ref{tableres} we show the mean age, burst strength,
metallicity and stellar mass and their corresponding standard
deviation values for all the star-forming regions studied. All those
clusters of solutions with probability higher than 20~per cent are
shown. This probability has been computed by dividing the number of
Monte Carlo particles within a given cluster relative the total number
of particles (10$^{3}$). Using the mean value for the highest
probability solution cluster of each star-forming region we obtained
the frequency histograms shown in Fig.~\ref{multhistos}.

The results shown in Table~\ref{tableres} indicate that 50~per cent of
the regions under study only show a cluster of solutions with
probability higher than 20~per cent. In the remaining regions the mean
differences obtained between the several solution clusters are
2.2\,Myr, 0.15\,dex and 0.14\,dex in age, burst strength and mass,
respectively, for models with a 15~per cent fraction of escaping
photons and any metallicity. These differences are even lower by using
other sets of models --1.8\,Myr, 0.10\,dex and 0.10\,dex,
respectively, for the subsolar metallicity models--. In any case,
these differences are significantly lower than the dispersion observed
in Fig.~\ref{multhistos}.

From Fig.~\ref{multhistos} we also deduce that there is no large
differences in the properties derived assuming 15~per cent or a null
fraction of Lyman photons escaping from the nebula. If we compare the
results obtained using subsolar metallicity models and those obtained
for the whole range in metallicity, it seems that a higher number of
regions older than 10\,Myr and with burst strength lower than 1~per
cent is obtained in the former case. Since the metallicity of the
ionized gas (see Sect.~\ref{starformingspectra}) is clearly lower than
solar, we are more confident with the results obtained using subsolar
metallicity models. The Principal Component Analysis performed on the
highest probability solution clusters indicate that the direction in
the ($t$,$b$,$Z$) space that better reproduces the data variance is
($u_t$,$u_b$,$u_Z$)=($+$0.707,$+$0.707,0.000). This fact suggests the
existence of a small degeneracy between age and burst strenght.

In the lower panels of Fig.~\ref{multhistos} we show the stellar
mass distribution. The stellar masses (see also Table~\ref{tableres})
have been computed using the $K$-band absolute magnitudes measured
within the apertures and the mean mass-to-light ratio of the
highest-probability solution cluster. These stellar masses were
corrected for the aperture effect by dividing them by the factors $f$
given in Table~\ref{tableres} (see Sect.~\ref{formingmethod}).

Finally, the age, burst strength and mass values obtained for these
regions are represented in Fig.~\ref{properties} using different
sized symbols. In Fig.~\ref{properties}a the size of the symbols
used is related with the age of the burst, larger symbols represent
younger regions. In Fig.~\ref{properties}b the symbol size is
proportional to the burst strength, and finally, in
Fig.~\ref{properties}c its size is proportional to the burst stellar
mass. Figs.~\ref{multhistos} and~\ref{properties}a show that the age
of the star-forming regions is well constrained between
5-13\,Myr. There is no significant age gradients across the different
structures observed in the H$\alpha$ image (see
Sect.~\ref{globalinterpretation}).

It should be noticed that the age and burst strength values derived
are not affected by the \object{Mrk~86} distance uncertainty, since
only aperture colors and equivalent widths have been used in this
work.

Since we have adopted in our models a fixed mass-to-ligth ratio and
colors for the underlying stellar population, the contamination from
intermediate aged populations could yield sistematically higher age,
burst strength and stellar mass values in some regions. This
could be the case of the \#45, \#49 and \#59 regions, contaminated
from the central starburst continuum emission. On the other hand, the
ignorance on the actual mass-to-light ratio of the underlying
population, as we pointed out in Sect.~\ref{under}, may introduce
slight uncertainties in the stellar masses derived. However, although
the absolute values for these masses would be quite uncertain, the
relative differences should be similar.

\subsection{Gas diagnostic for the star-forming regions}
\label{starformingspectra}

In Table~5 of Paper~I we gave the emission-line fluxes measured in
4.30$\times$2.65\,arcsec$^{2}$ regions centred in the maximum of the
emission knot section covered by the slit. 

The gas electron densities have been obtained from the
[\ion{S}{ii}]$\lambda$6716\AA/[\ion{S}{ii}]$\lambda$6731\AA\ line
ratio following Osterbrook (\cite{osterbrock}). When the latter line
ratio was not measurable we adopted an standard density of
$n_{\mathrm{e}}$=100\,cm$^{-3}$. In the cases where the
([\ion{O}{iii}]$\lambda$4959\AA$+$[\ion{O}{iii}]$\lambda$5007\AA)/[\ion{O}{iii}]$\lambda$4363\AA\
line ratio was measurable, we determined the electron temperatures
applying the algorithms given by Gallego (\cite{gallego95}). The
corresponding oxygen, nitrogen and helium abundances were also
computed using the algorithms given by Gallego
(\cite{gallego95}). Since the \ion{He}{ii}$\lambda$4686\AA\
emission-line fluxes were not measurable we could not determine the
\ion{He}{iii} abundance. In addition, since the 
[\ion{S}{iii}]$\lambda\lambda$9069,9532\AA\ emission lines were not 
accessible, we assumed a ionization correction factor of 1 and,
consequently, $y$=$y^{+}$. A summary of the electron densities,
temperatures and chemical abundances deduced is given in
Table~\ref{zs}.

Finally, we have compared the line ratios measured with the
predictions of a grid of model nebulae taken from Martin
(\cite{martin97}) and originally calculated with {\sc cloudy}. The
nitrogen-to-oxygen and carbon-to-oxygen abundance ratios used were
those employed by Martin (\cite{martin97}). The oxygen abundance was
0.2\,(O/H)$_{\sun}$.

\begin{figure*}
\resizebox{\hsize}{!}{\includegraphics{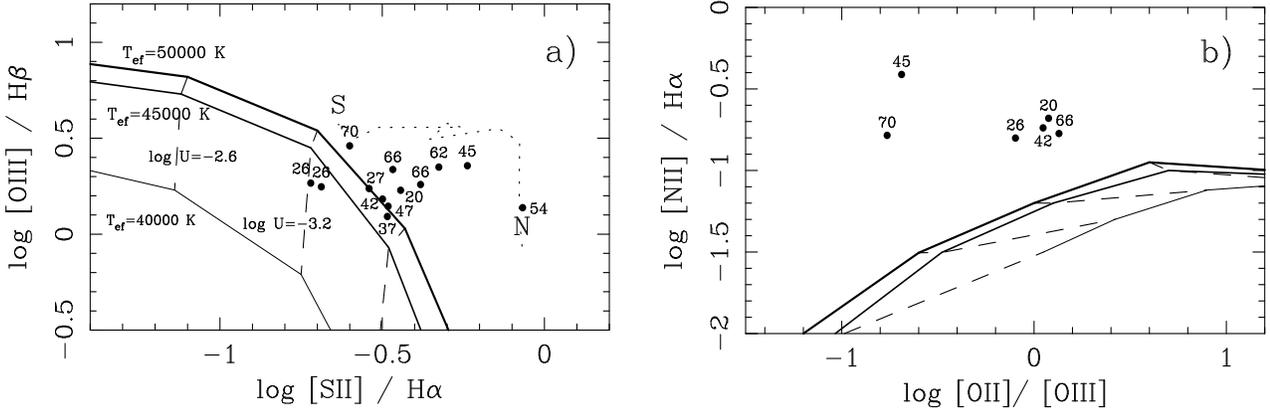}} 
\caption{Extinction corrected emission-line ratios. Only those regions 
with measurable Balmer decrements are shown. Photoionization models 
computed with {\sc cloudy} are drawn for 50000, 45000 and 40000\,K
effective temperatures and ionization parameters between
log\,U=$-$1.90 and $-$4.60. {\it Solid-lines} connect line ratios
computed with different ionization parameters and fixed effective
temperature. {\it Thicker-lines} mean higher temperatures. {\it
Dashed-lines} connect predictions for different effective temperatures
and fixed ionization parameter.}
\label{lineratios}
\end{figure*}

In Fig.~\ref{lineratios} the extinction corrected
[\ion{O}{iii}]$\lambda$5007\AA/H$\beta$,
[\ion{S}{ii}]$\lambda\lambda$6717,6731\AA/H$\alpha$,
[\ion{O}{ii}]$\lambda$3727\AA/[\ion{O}{iii}]$\lambda$5007\AA\ and
[\ion{N}{ii}]$\lambda$6583\AA/H$\alpha$ line ratios jointly with the
models predictions have been plotted. We have drawn models with
effective temperatures in the range 40000-50000\,K and ionization
parameters between log\,U=$-$1.91 and log\,U=$-$4.60. {\it
Solid-lines} represent the change in the line ratios for different
ionization parameter values between log\,U=$-$1.91 and $-$4.60 at
increments of 4.7 in U. Thicker lines mean higher temperatures. The
{\it dashed-lines} represent the change in the line ratios as a
function of the effective temperature for a fixed ionization
parameter. In all these diagrams the ionization parameter of the
models increases from right to left. In Fig.~\ref{lineratios}a we
also show the change in the line ratios measured along the major axis
of the expanding bubble
\object{Mrk~86}-B (GZG) from North to South ({\it dotted-line}). 

\begin{figure*}
\resizebox{\hsize}{!}{\includegraphics{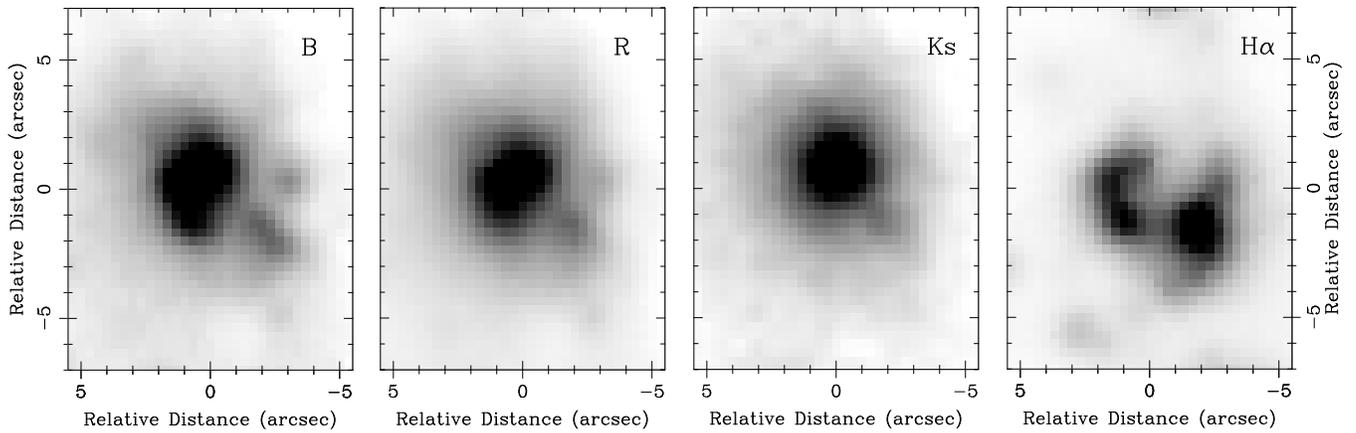}} 
\caption{\#26 and \#27 regions. $B$, $R$, $Ks$ and H$\alpha$ images are shown. North is up and East is to the left. The scale is 0\farcs33\,pixel$^{-1}$. The size of these images is approximately 0.4\,kpc$\times$0.5\,kpc.}
\label{knot26}
\end{figure*}

As it was pointed out by Martin (\cite{martin97}), the bulk of the
discrepancy of these line ratios with the prediction of
photoionization models suggests the existence of an aditional
excitation mechanism. This discrepancy will be higher using lower
metallicity models. The contribution of this additional mechanism (or
mechanisms) is more significant, relative to that produced by
photoionization, in the case of the \#45,
\#54 and \#70 star-forming regions. In the latter case, the anomalous
line-ratios measured are probably related with enhanced shocked gas
emission in the \object{Mrk~86}-B bubble fronts (see GZG). The
contamination from the \object{Mrk~86}-B north lobe could be also
responsible for the line ratios measured in the \#54 region.

\subsection{Comments on several individual regions}
\label{comments}

\noindent
{\it \#9, \#10, \#12, \#22, \#55, \#57, \#79, \#84 and \#85}: All
these regions show photometric H$\alpha$ emission, but very faint or
undetectable $R$-band continuum emission. There are two feasible
explanations for this very faint continuum emission.

First, these regions could be high gas density clumps photoionized by
distant stellar clusters. Then, they should be placed in regions with
intense diffuse H$\alpha$ emission. This could be the case of the \#9,
\#10, \#12, \#22, \#55 and \#57 regions. 

On the other hand, at the early evolutionary stages of a starburst the
emission-line equivalent widths can be as high as 1000\AA. Thus,
star-forming regions with low burst strength will be only detectable
by their H$\alpha$ or [\ion{O}{iii}]$\lambda$5007\AA\ emission. This
could be the case of the \#79, \#84 and \#85 regions.

\noindent
{\it \#26 \& \#27}: These regions conform a massive association (see
GZG) with very complex structure. The best-fitting model for the
\#26 region yields an age of about $\sim$10\,Myr with a high burst
strength value. On the other hand, the \#27 region is a younger burst
(5\,Myr) with low burst strength and complex H$\alpha$ emission
structure (see Fig.~\ref{knot26}). The peculiar velocity profile
obtained by GZG suggests that this association could belong to an
independent stellar system merged with \object{Mrk~86}. In
Fig.~\ref{knot26} we show the continuum and H$\alpha$ structure of
this association. From this figure is not clear if the H$\alpha$
emission arises from photoionization of the \#26, \#27 or both stellar
clusters. The H$\alpha$ fluxes of the \#26 and \#27 regions given in
Table~5 of Paper~I are those measured for the north-east and
south-west structures shown in Fig.~\ref{knot26} ({\it right
panel}), respectively.

\noindent
{\it \#42, \#70 \& \#18}: These regions correspond to the starburst
precursors of the \object{Mrk~86}-A, \object{Mrk~86}-B and
\object{Mrk~86}-C expanding bubbles, respectively (see Martin 
\cite{martin}, GZG). From Fig.~\ref{properties} we observe that these
regions have extreme properties. In particular, the \#70 region shows
the highest burst strength (excepting \#45 region and the central
starburst) and the \#18 region is the youngest of the regions analyzed
in Sect.~\ref{starforming}.

\section{Global interpretation of the \object{Mrk~86} star formation history}
\label{globalinterpretation}

Summarizing, we have derived the properties of three well defined
stellar populations;
\begin{enumerate}
\item {\bf Underlying population}: Exponential surface profile, red colors, 
no significant color gradients. The ages derived range between 5 and
13\,Gyr.
\item {\bf Central starburst}: 20~per cent burst strength, 30\,Myr old 
and total stellar mass of about 9$\times$10$^{6}$\,M$_{\sun}$.
\item {\bf Star-forming regions}: Low burst strengths ($\sim$2~per cent), 
ages between 5-13\,Myr and low gas metallicities. No significant age
or metallicity gradients are observed.
\end{enumerate}

\subsection{What has been ocurring during the last 30\,Myr?}
\label{globalrecent}
Most of the \ion{H}{ii} regions in \object{Mrk~86} are located in an
bar oriented in north-south direction and placed 20\arcsec\ west of
the galactic center (see Fig.~\ref{properties} and Fig.~2 of
Paper~I) and in an arc going from south-east (\#58 \&
\#64 regions) to north (\#13 \& \#16 regions) of the central
starburst. There is no significant age gradients across these
structures, suggesting the existence of a large-scale triggering
mechanism that activated the star formation simultaneously in most of
these regions about 10\,Myr ago.

We propose that the large-scale triggering mechanism for the recent
star formation was produced by the growth of a superbubble originated
at the galaxy central starburst by the energy deposition of stellar
winds and supernova explosions. In some cases, local triggering
mechanisms are also present. This could be the case of the complexes
formed by the \#26-\#27, \#26-\#18, \#58-\#64 and
\#16-\#13 regions. These complexes are constituted by an {\it evolved}
burst (age$\geq$10\,Myr) --the former--, and a very recent
star-forming event, about 5\,Myr old --the latter-- (see Stewart et
al. \cite{stewart})

Hereafter, we will mainly focus on the large-scale triggering
mechanism described above. Following the models of Silich \&
Tenorio-Tagle (\cite{silich}, STT hereafter; see also De Young \&
Heckman \cite{young94}), these collective supernova remnants grow at
first elongated in the direction perpendicular to the galaxy disk. Due
to this, the remnant blows out into the galaxy halo, leading to the
formation of a secondary shell of swept out gas. About 20\,Myr after
the first supernova explosions (for the STT A100 model) the
superbubble surrounds the inner densest part of the disk. Then, the
leading shock, going through outer less dense regions, merges with its
symmetrical counterpart. During this merging process a large fraction
of swept out mass is strangled shaping a dense toroid. This region
does not participate in the general outward motion, being strongly
compressed towards the galaxy plane. As a result of this compression
the activation of the star formation in the gas toroid could be
produced if the mean surface density will be higher than
5-10\,M$_{\sun}$\,pc$^{-2}$ (Skillman et al. \cite{skillman};
Kennicutt \cite{kenni89}; see also Taylor et al. \cite{taylor94} for a
sample of \ion{H}{ii} galaxies).

The predictions described above correspond to the A100 model of
STT. This model assumes a 10$^{10}$\,M$_{\sun}$ massive galaxy with
10~per cent gas content, ISM central density of
$n_{0}$=20.2\,cm$^{-3}$, burst energy of
$E_{\mathrm{burst}}$=10$^{56}$\,erg and gas metallicity,
Z=0.3\,Z$_{\sun}$. For this model, adopting a toroid width of 500\,pc
and a strangled mass of 2$\times$10$^{8}$\,M$_{\sun}$, the surface
density of the gas at the toroid will be about
64\,M$_{\sun}$\,pc$^{-2}$. This value is intermediate between the gas
surface densities measured in normal disks and in the infrared-seleted
starbursts studied by Kennicutt (\cite{kenni98}). The
galactocentric distance and epoch predicted for the formation of the
toroid in the A100 STT model are 1.5\,kpc and 25-30\,Myr,
respectively.

In the case of \object{Mrk~86} the epoch for the formation of this
toroid should be about 20\,Myr (the age difference between the
starburst and currently star-forming regions) at a radius of
0.8\,kpc. Although these values well agree with the properties
deduced for the A100 STT model, the high dependence of the evolution
of the superbubble on the ISM density profile and the distance
uncertainty for \object{Mrk~86} prevent us to carry out a more
quantitative analysis. However, this scenario provides a reliable and
attractive explanation for the evolution of the star formation
activity in \object{Mrk~86}.

\begin{figure*}
%\resizebox{8.cm}{!}{\includegraphics{fig6a.eps}} 
%\resizebox{8.3cm}{!}{\includegraphics{fig6b.eps}} 
\vspace{8.5cm}
\caption{{\it Left-panel}: VLA (D configuration) 21cm line intensity 
map courtesy of E. Wilcots. North is up and East is to the left. The
edge-on spiral galaxy east-south of \object{Mrk~86} is
\object{UGC~4278}. The grey scale flux ranges between 0.025 and 3
kJy\,m\,s$^{-1}$\,beam$^{-1}$ (see the upper bar on the figure). The
beam size for this configuration is 46\arcsec (HPBW). {\it
Right-panel}: $R$-band image obtained with the Wide-Field Camera (WFC)
at the INT (see Paper~I).}
\label{wilcotsfig}
\end{figure*}

Finally, it should be noticed that the time since the formation of the
toroid ($\sim$10\,Myr) is significantly lower than the galaxy rotation
period ($\sim$90\,Myr), avoiding the disruption of this toroid by
differential rotation. The galaxy rotation period was obtained using a
projected angular velocity of 44\,km\,s$^{-1}$\,kpc$^{-1}$ and
assuming an inclination of the rotation axis relative to the plane of
the sky of 50$^{\circ}$ (GZG; see also Gil de Paz \cite{gilphd}).

The scenario described above is similar to that observed in other
nearby dwarf star-forming galaxies. Thus, in the \object{LMC} the two
arcs of stellar clusters place on the LMC4 superbubble rim round a
population of 30\,Myr old supergiants and Cepheid variables (Efremov
\& Elmegreen \cite{efremov}). A similar study for the DEM192 superbubble also in 
the \object{LMC} was carried out by Oey \& Smedley (\cite{oey}).

\subsection{What did happen 30\,Myr ago?}
\label{globalintermediate}
Although we can accept that most of the recent star-forming activity
in \object{Mrk~86} was originated by a superbubble produced at the
central starburst, we need to solve the question about the nature of
the triggering mechanism of the star formation in the galaxy central
starburst.

Under the evolutionary scenario described above, the observational
properties of \object{Mrk~86} about 30\,Myr ago would be very similar
to those of the nucleated Blue Compact Dwarfs (nE BCDs). It should
show a very bright central starburst superimposed on an extended
underlying component (see Papaderos et al. \cite{papaI} and references
therein). This fact suggests the existence of an evolutionary
connection between the different kinds of Blue Compact Dwarf
galaxies. Therefore, the question about the star formation triggering
in the central starburst of \object{Mrk~86} is equivalent to the
problem of the activation of the star formation in the BCD galaxy
population as a whole.

Taylor et al. (\cite{taylor93}) hypothesized that the close passage of
a companion galaxy has triggered the present burst of star formation
(the central starburst in the \object{Mrk~86} case) in many (if not
all) of the BCD galaxies. These kind of distant encounters are well
acepted to lead the lost of angular momentum in the gas component,
resulting in the fall of large amounts of gas to the galaxy central
regions (Barnes \& Hernquist \cite{barnes92}; Taylor et
al. \cite{taylor94}; Mihos \& Hernquist \cite{mihos96}). This radial
inflow will be very efficient in \object{Mrk~86}, because of the low
dark matter content expected for this object (GZG; see Taylor
\cite{taylor97}).

The first candidate for such a tidal interaction is
\object{NGC~2537A} ($\alpha$(2000)=08$^{h}$13$^{m}$40.9$^{s}$
$\delta$(2000)=$+$45\degr 59\arcmin 41\arcsec). However, the study of
Heyd \& Wiyckoff (\cite{heyd}) and the 21cm line intensity map (see
Fig.~\ref{wilcotsfig}) show that this object is clearly a background
galaxy. Then, the next candicate for the central starburst triggering
is \object{UGC~4278} ($\alpha$(2000)=08$^{h}$13$^{m}$58.8$^{s}$
$\delta$(2000)=$+$45\degr 44\arcmin 36\arcsec), a nearly edge-on
spiral galaxy placed at a projected distance of about 33\,kpc relative
to \object{Mrk~86} (see Fig.~\ref{wilcotsfig}). Its heliocentric
velocity is 553\,km\,s$^{-1}$ (Goad \& Roberts
\cite{goad}; see also Schneider \& Salpeter \cite{schneider}),
and its radial velocity relative to the CMB is 715\,km\,s$^{-1}$.

The offset of this galaxy relative to the Tully-Fisher template
obtained by Giovanelli et al. (\cite{giovanelli97}) is
0.66$^{\mathrm{m}}$ in the $I$-band (R. Giovanelli,
priv. comm.). Therefore, its peculiar radial velocity will be
$-$253\,km\,s$^{-1}$, being the radial velocity corrected for peculiar
motions 968\,km\,s$^{-1}$. Then, the distance to
\object{UGC~4278} could range between 13 and 19\,Mpc, respectively for
$h$=0.75 and 0.5. The large difference between this value and that
derived by Sharina et al. (\cite{sharina99}; see also Sect.~2.1 of
Paper~I) for \object{Mrk~86} suggests that \object{UGC~4278} is also a
background galaxy.

Therefore, another triggering mechanism should be argued to
explain the activation of the star formation in the \object{Mrk~86}
central starburst. This triggering mechanism could be related with the
presence of previous massive star forming events. Our study has not
revealed the existence of such an intermediate aged stellar
population, probably excepting the \#26 and \#27 regions. As we
indicated in Sect.~\ref{comments} (see also GZG), the association
constituted by the \#26 and \#27 regions shows a very steep velocity
gradient that could be produced by the velocity field of an
independent low mass system merged or in process of merging with
\object{Mrk~86}. In that case, this merging process could be
responsible for radial inflow of gas that led to the star formation
activation in the central regions of \object{Mrk~86}. However, the
large-scale distribution of neutral hydrogen in this galaxy (see
Fig.~\ref{wilcotsfig}; E. Wilcots, priv. comm.) indicates that, if
this merging process took place, it should occur long time ago,
probably several orbital periods ago.

\section{Summary and conclusions}
\label{conclusions}

The main conclusions derived from this work are summarized hereafter.

\begin{itemize}

\item
The Blue Compact Dwarf galaxy \object{Mrk~86} is constituted by three
well defined stellar populations. An evolved (5-13\,Gyr old) stellar
component characterized by an exponential light profile, no
significant color gradients and low metallicity. A massive
($\sim$9$\times$10$^{6}$\,M$_{\sun}$) central starburst, about 30\,Myr
old, with very low dust content and high burst strength ($\sim$20~per
cent). And, finally, a young stellar population distributed in, at
least, 46 star-forming regions. These star-forming regions are
characterized by very low metallicities, burst strengths and
stellar masses.

\item 
The distribution of the star-forming regions properties suggest that
their star formation triggering is related with a large-scale
mechanism. Following the models of Silich \& Tenorio-Tagle
(\cite{silich}), we propose that the growth of a superbubble produced
by the energy deposition at the galaxy central starburst led to the
formation of a dense toroid of ISM mass. Then, the high gas surface
densities reached produced the activation of the current star-forming
activity.

\item 
Finally, we studied the possible triggering mechanisms for the
activation of the star formation in the galaxy central
starburst. Since both candidates for a distant encounter,
\object{NGC~2537A} and \object{UGC~4278}, seem to be background 
galaxies, the merging with a low mass companion seems the most
feasible explanation for this central starburst activation.

\end{itemize}

\section*{Acknowledgments}
    Based on observations with the JKT, INT and WHT operated on the
    island of La Palma by the Royal Greenwich Observatory in the
    Spanish Observatorio del Roque de los Muchachos of the Instituto
    Astrof\'{\i}sico de Canarias. Based also on observations collected
    at the German-Spanish Astronomical Center, Calar Alto, Spain,
    operated by the Max-Planck-Institut f\"{u}r Astronomie (MPIA),
    Heidelberg, jointly with the spanish 'Comisi\'on Nacional de
    Astronom\'{\i}a'. This research has made use of the NASA/IPAC
    Extragalactic Database (NED) which is operated by the Jet
    Propulsion Laboratory, California Institute of Technology, under
    contract with the National Aeronautics and Space Administration.
                   
    We are grateful to Carme Jordi and D. Galad\'{\i} for obtaining
    the $V$-band image. We would like to thank C. S\'anchez Contreras
    and L.F. Miranda for obtaining the high resolution spectra. We
    also thank A. Alonso-Herrero for her help in the acquisition and
    reduction of the near-infrared images. We also acknowledge the
    referee Dr. Tosi for several helpful comments. We thank to
    C.E. Garc\'{\i}a Dab\'o, J. Cenarro, M.E. Sharina, S.A. Silich and
    J. Gorgas for stimulating conversations. Finally, we are very
    grateful to R. Giovanelli for his kind help on analyzing the
    properties of \object{UGC~4278}. This research has been supported
    in part by the grants PB93-456 and PB96-0610 from the Spanish
    'Programa Sectorial de Promoci\'on del Conocimiento'. A. Gil de
    Paz acknowledges the receipt of a 'Formaci\'on del Profesorado
    Universitario' fellowship from the spanish 'Ministerio de
    Educaci\'on y Cultura'.


\begin{thebibliography}{}
\bibitem[1999]{aloisi99}
Aloisi A., Tosi M., Greggio L., 1999, AJ 118, 302 
\bibitem[1966]{arp66}
Arp H., 1966, Atlas of Peculiar Galaxies. California Institute of
Technology, Pasadena
\bibitem[1992]{barnes92}
Barnes J.E., Hernquist L.E., 1992, ARA\&A 30, 705 
\bibitem[1983]{bruzual}
Bruzual A.G., 1983, ApJ 273, 105 
\bibitem[1989]{burkert}
Burkert A., 1989, Ph.D. Thesis, Munich
\bibitem[1982]{burstein}
Burstein D., Heiles C., 1982, AJ 87, 1165
\bibitem[1996]{calzetti}
Calzetti D., Kinney A.L., Storchi-Bergmann T., 1996, ApJ 458, 132
\bibitem[1984]{campbell}
Campbell A.W., Terlevich R., 1984, MNRAS 211, 15
\bibitem[1994]{young94}
De Young D.S., Heckman T.M., 1994, ApJ 431, 598
\bibitem[1998]{doublier98}
Doublier V., 1998, Ph.D. Thesis, Marsella
\bibitem[1991]{drinkwater}
Drinkwater M., Hardy E., 1991, AJ 101, 94  
\bibitem[1998]{efremov}
Efremov Y.N., Elmegreen B.G., 1998, MNRAS 299, 643
\bibitem[1988]{fanelli}
Fanelli M.N., O'Conell R.W., Thuan T.X., 1988, ApJ 334, 665
\bibitem[1995]{gallego95}
Gallego J., 1995, Ph.D. Thesis, Universidad Complutense de Madrid
\bibitem[2000]{gilphd}
Gil de Paz A., 2000, Ph.D. Thesis, Universidad Complutense de Madrid 
\bibitem[1999]{GZG}
Gil de Paz A., Zamorano J., Gallego J., 1999, MNRAS 306, 975 (GZG) 
\bibitem[2000a]{paperI}
Gil de Paz A., Zamorano J., Gallego J., Dom\'{\i}nguez F. de B., 2000a,
A\&A (Paper~I)
\bibitem[2000b]{nirucm}
Gil de Paz A., Arag\'on-Salamanca A., Gallego J., et al., 2000b, MNRAS 316, 357
\bibitem[1997]{giovanelli97}
Giovanelli R., Haynes M.P., Herter T., et al., 1997, AJ 113, 53
\bibitem[1981]{goad}
Goad J.W, Roberts M.S., 1981, ApJ 250, 79
\bibitem[1993]{gorgas93}
Gorgas J., Faber S.M., Burstein D., et al., 1993, ApJS 86, 153
\bibitem[1999]{gorgas}
Gorgas J., Cardiel N., Pedraz S., Gonz\'alez J., 1999, A\&AS 139, 29
\bibitem[1992]{heyd}
Heyd R., Wyckoff S., 1992, BAAS 181, \#45.09
\bibitem[1990]{hoffman}
Hoffman G.L., Salpeter E.E., Helou G., 1990, in the proceedings of the Edwin Hubble Centennial
Symposium, 67
\bibitem[1989]{kenni89}
Kennicutt R.C., 1989, ApJ 344, 685
\bibitem[1998]{kenni98}
Kennicutt R.C., 1998, ApJ 498, 541 
\bibitem[1986]{kunth86}
Kunth D., Sargent W.L.W., 1986, ApJ 300, 496
\bibitem[1988]{kunth88}
Kunth D., Maurogordato S., Vigroux L., 1988, A\&A 204, 10
\bibitem[1985]{loose85}
Loose H.-H., Thuan T.X., 1985, The Morphology and Structure of BCDGs
from CCD Observations, in: Kunth, D., Thuan, T.X., and Van,
J.T.T.(eds.) Star-Forming Dwarf Galaxies. Editions Fronti\`{e}res.
\bibitem[1969]{markarian69}
Markarian B.E., 1969, Astrofizika, 5, 443
\bibitem[1995]{marlowe95}
Marlowe A.T., Heckman T.M., Wyse R.F.G., Schommer R., 1995, 438, 563
\bibitem[1997]{martin97}
Martin C.L., 1997, ApJ 491, 561 
\bibitem[1998]{martin}
Martin C.L., 1998, ApJ 506, 222
\bibitem[1996]{mihos96}
Mihos J.C., Hernquist L., 1996, ApJ 464, 641
\bibitem[1976]{morrison}
Morrison D.F., 1976, Multivariate Statistical Methods, McGraw-Hill
Book Co., Singapore
\bibitem[1987]{murtagh}
Murtagh F., Heck A., 1987, Multivariate Data Analysis, D. Reidel
Publishing Co., Dordrecht, Holland
\bibitem[1997]{norton}
Norton S., Salzer J.J., 1997, BAAS 190
\bibitem[1998]{oey}
Oey M.S., Smedley S.A., 1998, AJ 116, 1263
\bibitem[1989]{osterbrock}
Osterbrock D.E., 1989, Astrophysics of Gaseous Nebulae and Active
Galactic Nuclei. University Science Books, Mill Valley, California.
\bibitem[1996a]{papaI}
Papaderos P., Loose H.-H., Thuan T.X., Fricke K.J., 1996a, A\&AS 120, 207
\bibitem[1996b]{papaII}
Papaderos P., Loose H.-H., Fricke K.J., Thuan T.X., 1996b, A\&A 314, 59 
\bibitem[1992]{schneider}
Schneider S.E., Salpeter E.E., 1992, ApJ 385, 32
\bibitem[1972]{searle72}
Searle L., Sargent W.L.W., 1972, ApJ 173, 25
\bibitem[1973]{searle73}
Searle L., Sargent W.L.W., Bagnuolo W.G., 1973, ApJ 179, 427
\bibitem[1932]{shapley32}
Shapley H., Ames A., 1932, Ann. Harvard College Obs. 88, No. 2
\bibitem[1999]{sharina99}
Sharina M.E., Karachentsev I.D., Tikhonov N.A., 1999, Astronomy Letters 25, 322
\bibitem[1998]{silich}
Silich S.A., Tenorio-Tagle G., 1998, MNRAS 299, 249 (STT)
\bibitem[1987]{silk87}
Silk J., Wyse R.F.G., Shields G.A., 1987, ApJ 322, L59
\bibitem[1987]{skillman}
Skillman E.D., Bothun G.D., Murray M.A., Warmels R.H., 1987, A\&A 185, 61
\bibitem[2000]{stewart}
Stewart S.G., Fanelli M.N., Byrd G.G., et al., 2000, ApJ 529, 201
\bibitem [1997]{taylor97}
Taylor C.L., 1997, ApJ 480, 524
\bibitem[1993]{taylor93}
Taylor C.L., Brinks E., Skillman E.D., 1993, AJ 105, 128
\bibitem[1994]{taylor94}
Taylor C.L., Brinks E., Pogge R.W., Skillman E.D., 1994, AJ 107, 971
\bibitem[1983]{thuan83}
Thuan T.X., 1983, ApJ 268, 667
\bibitem[1991]{thuan91}
Thuan T.X., 1991, Observations and Models of Blue Compact Dwarf
Galaxies, in: Leitherer C., Walborn R.N., Heckman T.M., Norman
C.A. (eds.) Massive Stars in Starbursts, Cambridge University Press,
p.183.
\bibitem[1981]{thuan81}
Thuan T.X., Martin G.E., 1981, ApJ 247, 823
\bibitem[1994]{worthey}
Worthey G., 1994, ApJS 95, 107

\end{thebibliography}
\end{document}